\documentclass[11pt,a4paper]{article}
\usepackage{jheppub,xcolor}
\pdfoutput=1
\usepackage[utf8]{inputenc}
\graphicspath{{./figs/}}

\usepackage{amssymb}
\usepackage{dcolumn}	
\usepackage{bm}			
\usepackage{bbm} 		
\usepackage{multirow}
\usepackage{slashed}
\usepackage{blkarray}
\usepackage{booktabs}
\usepackage{makecell}
\usepackage{tikz}
\usepackage[compat=1.1.0]{tikz-feynman}
\usepackage{longtable}
\usepackage{placeins}
\usepackage{listings}
\usepackage{amsmath}
\usepackage{xcolor}
\usepackage{pythonhighlight}
\usepackage{subcaption} 
\usepackage{rotating}
\usepackage{float}      
\usepackage{adjustbox}  
\usepackage{afterpage}
\usepackage{soul}

\def\gsim{\mathrel{\rlap{\lower4pt\hbox{\hskip1pt$\sim$}}
    \raise1pt\hbox{$>$}}}         
\def\lsim{\mathrel{\rlap{\lower4pt\hbox{\hskip1pt$\sim$}}
    \raise1pt\hbox{$<$}}}         

\definecolor{codegreen}{rgb}{0,0.6,0}
\definecolor{codegray}{rgb}{0.5,0.5,0.5}
\definecolor{codepurple}{rgb}{0.58,0,0.82}
\definecolor{backcolour}{rgb}{0.95,0.95,0.92}

\lstdefinestyle{mystyle}{
    backgroundcolor=\color{backcolour},   
    commentstyle=\color{codegreen},
    keywordstyle=\color{magenta},
    numberstyle=\tiny\color{codegray},
    stringstyle=\color{codepurple},
    basicstyle=\ttfamily\footnotesize,
    breakatwhitespace=false,         
    breaklines=true,                 
    captionpos=b,                    
    keepspaces=true,                 
    numbers=left,                    
    numbersep=5pt,                  
    showspaces=false,                
    showstringspaces=false,
    showtabs=false,                  
    tabsize=2
}
\definecolor{Red}{rgb}{1.,0.,0.}

\newcommand{\lag}{\mathcal{L}}

\newcommand{\mcO}{\mathcal{O}}

\newcommand{\sss}{\scriptscriptstyle}
\newcommand{\pdp}{\ensuremath{\varphi^\dagger\varphi}}
\newcommand{\ccc}[3]{c_{#2}^{#1 (#3)}}
\newcommand{\qq}[3]{\ensuremath{\mathcal{O}_{#2}^{#1 (#3)}}}

\usepackage[normalem]{ulem}

\newcommand{\bc}{\begin{center}}
\newcommand{\ec}{\end{center}}
\newcommand{\ba}{\begin{array}}
\newcommand{\ea}{\end{array}}

\newcommand{\be}{\begin{equation}}
\newcommand{\ee}{\end{equation}}
\newcommand{\bea}{\begin{align}}
\newcommand{\eea}{\end{align}}

\usepackage{tabularx}

 \newcolumntype{C}[1]{>{\centering\arraybackslash}p{#1}}

\usepackage[normalem]{ulem}

 \def\lag{\ensuremath{\mathcal{L}}}
 \def\lra#1{\overset{\text{\scriptsize$\leftrightarrow$}}{#1}}
\newcommand{\smefit}{{\sc \small SMEFiT}}
\newcommand{\dsix}{{\sc \small DsixTools}}
\newcommand{\sold}{{\tt SOLD}}
\newcommand{\mmeft}{{\tt Matchmakereft}}

\numberwithin{equation}{section}
\numberwithin{figure}{section}
\numberwithin{table}{section}

\title{The effect of the two-loop SMEFT RGEs at future colliders}
\author[a]{Luca~Mantani,}
\author[b]{Pablo~Olgoso,}
\author[b]{and Alejo~N.~Rossia}

\affiliation[a]{Instituto de F\'isica Corpuscular (IFIC), Universidad de Valencia-CSIC, E-46980 Valencia, Spain}
\affiliation[b]{Dipartimento di Fisica e Astronomia ``G. Galilei'', Universit\`a di Padova, and Istituto Nazionale di Fisica Nucleare, Sezione di Padova, Via F. Marzolo 8, I-35131, Padova, Italy.}

\emailAdd{luca.mantani@uv.es}
\emailAdd{pablo.olgosoruiz@unipd.it}
\emailAdd{alejonahuel.rossia@unipd.it}

\date{\today}

\abstract{The search for New Physics requires ever increasing precision from experimental and theoretical efforts. Within the Standard Model Effective Field Theory (SMEFT) framework, the latest achievement in this quest has been the complete computation of the two-loop Renormalisation Group Equations (RGEs) for the Wilson Coefficients of dimension-six operators. In this work, we solve the two-loop SMEFT RGEs with full numerical integration and compare the evolution matrix obtained at one and two loops to analyze how two-loop contributions alter mixing patterns and break zeroes present at one-loop order.
Then, we perform a first comprehensive analysis of the impact of the two-loop RGEs in phenomenological studies at HL-LHC and FCC-ee. From a bottom-up perspective, we carry out individual and global fits at linear and quadratic level for a set of 61 Wilson coefficients and compare against the results obtained by including only one-loop RGE effects. We find non-negligible two-loop induced effects in some cases, in particular for four-quark, top Yukawa and Higgs-gluon operators. 
From a top-down perspective, we perform fits to all the scalar and fermion extensions of the Granada dictionary matched onto SMEFT at one-loop level, including for the first time the couplings that enter only at one loop, and find percent-level effects in the sensitivity to the couplings of some models.
}

\keywords{Standard Model Effective Field Theory, Renormalisation Group Equations, Future Particle Colliders}

\begin{document}

\maketitle

\section{Introduction}
\label{sec:introduction}

The search for physics Beyond the Standard Model (BSM) has broadened significantly 
over the last decade. Direct searches for new particles are now complemented by 
indirect searches: the precise scrutiny of Standard Model (SM) predictions for 
deviations that would signal the presence of New Physics (NP). This dual strategy 
is expected not only to continue but to intensify as experimental precision improves.

Effective Field Theories (EFTs) provide the natural framework for such indirect 
searches, and can be approached from two directions. One can work bottom-up, 
parametrizing deviations from SM predictions agnostically through the coefficients 
of higher-dimensional operators, without committing to a specific BSM model. 
Alternatively, a top-down approach interprets those coefficients in terms of the 
parameters of an underlying UV theory, forging a direct connection between 
low-energy measurements and high-energy dynamics. Both perspectives are essential 
to extract the maximum information from experimental data, and both have seen 
significant progress in recent years.

A key driver of this progress has been extending analyses beyond leading order. 
In particular, operator mixing and renormalization group running are essential ingredients for correctly predicting low-energy 
phenomenology from high-scale Wilson coefficients. The one-loop Renormalization 
Group Equations (RGEs) for SMEFT have been known for over a decade 
\cite{Jenkins:2013zja,Jenkins:2013wua,Alonso:2013hga} and are now fully 
automated \cite{Aebischer:2018bkb,Fuentes-Martin:2020zaz,DiNoi:2022ejg}. Their 
impact on the interpretation of electroweak-scale and flavor measurements is 
well established~\cite{deBlas:2015aea,Greljo:2023bdy,Allwicher:2023shc}. It is 
now equally clear that one-loop RGEs are indispensable for the interpretation of 
high-energy collider measurements: as LHC precision has grown, and with 
order-of-magnitude improvements expected at the HL-LHC and future colliders, a 
series of studies has demonstrated their importance in any SMEFT interpretation of 
collider data~\cite{Battaglia:2021nys,Aoude:2022aro,DiNoi:2023onw,Garosi:2023yxg,
Maltoni:2024dpn,Greljo:2023bdy}. One-loop RGEs have since been incorporated 
into state-of-the-art global SMEFT fits~\cite{Bartocci:2024fmm,terHoeve:2025gey,
deBlas:2025xhe}, where they have again been shown to improve both the precision 
and accuracy of collider measurement interpretations.

On the top-down side, the development of automated one-loop matching 
tools~\cite{DasBakshi:2018vni,Carmona:2021xtq,Fuentes-Martin:2022jrf,
Guedes:2023azv,Guedes:2024vuf} has made it possible to systematically constrain 
the effects of any weakly-coupled SM extension.
Among all possibilities, the models 
that contribute at leading order, i.e. tree level and dimension six, are those catalogued in the Granada 
dictionary~\cite{deBlas:2017xtg,delAguila:2000rc,delAguila:2008pw,
delAguila:2010mx,deBlas:2014mba}. The interplay between these extensions and 
SMEFT RGEs has proven particularly significant: it has been shown that RGE effects 
are crucial to constrain them using the high-precision $Z$-pole observables 
foreseen in the FCC programme~\cite{Allwicher:2024sso}.

As experimental precision continues to improve, the assessment of the importance of two-loop RGE 
effects becomes increasingly pressing. 
While partial results have appeared in the 
literature~\cite{Aebischer:2022anv,Born:2024mgz,DiNoi:2024ajj,Ibarra:2024tpt,
Duhr:2025zqw,DiNoi:2025arz,Haisch:2025lvd,Haisch:2025vqj,Duhr:2025yor,
DiNoi:2025tka,Banik:2025wpi,Henriksson:2025vyi,Zhang:2025ywe,Duhr:2026btw}, the complete 
two-loop SMEFT dimension-6 RGEs have only recently become available~\cite{
Born:2026xkr}. The relevance of this result is not merely formal; the same 
precision frontier that made one-loop RGEs indispensable, and the even higher 
precision expected at the FCC, implies that two-loop corrections to operator 
running and mixing may be non-negligible in global SMEFT fits.

The scope of this paper is to give a first quantitative answer to this question, 
both from the bottom-up perspective and for the models in the Granada dictionary, since they are the ones with leading two-loop RGE effects. We 
stress that a fully consistent two-loop analysis would require two-loop predictions 
for all observables and two-loop matching conditions for the UV models, 
milestones that remain in the future. The present study should therefore be viewed 
as a motivation and guide for the road ahead: a preliminary assessment of 
phenomenological impact designed to inform where two-loop corrections matter most.

The paper is organized as follows. In Section~\ref{sec:solvingRGEs} we discuss the 
solution of the two-loop dimension-6 SMEFT RGEs and compare with the one-loop 
result. Section~\ref{sec:smeft_fits} presents bottom-up individual and global SMEFT 
fits to HL-LHC and FCC-ee projections, quantifying the impact of two-loop versus 
one-loop RGEs. In Section~\ref{sec:uv_fits} we adopt the top-down perspective, 
performing fits to the UV couplings of one-loop-matched scalars and fermions from 
the Granada dictionary. Section~\ref{sec:conclusions} summarizes our findings and 
outlines future directions. Three appendices provide supporting material: 
Appendix~\ref{app:smefit_basis} details the SMEFT basis conventions used 
throughout, Appendix~\ref{app:red_fit} illustrates how the two-loop RGE affects global fits, and Appendix~\ref{app:UV_details} gives details on the UV models 
discussed in Section~\ref{sec:uv_fits}.

\section{Solution of the 2-loop RGEs}
\label{sec:solvingRGEs}

In this section we describe how we obtained and implemented the two-loop SMEFT 
RGE matrix, and assess the structural differences with respect to the one-loop 
result. We first detail the basis translation and numerical implementation 
(Section~\ref{sec:basis_translation}), and then compare the one- and two-loop 
evolution matrices to identify the new operator mixings and the size of corrections 
to existing ones (Section~\ref{sec:comparison_1_vs_2_loops}).

\subsection{Basis translation and implementation}
\label{sec:basis_translation}

The two-loop SMEFT RGE matrix was obtained from~\cite{Born:2026xkr}, where it is 
provided in the Mainz basis, and translated to the Warsaw basis following the 
prescription therein. We validated the basis rotation by comparing the result at one loop 
against the \dsix~implementation and the literature~\cite{Jenkins:2013wua,Jenkins:2013zja,Alonso:2013hga,Aoude:2022aro,Maltoni:2024dpn}.
We then implemented the two-loop SMEFT RGE matrix in the Warsaw basis in a custom modified version of 
\dsix~\cite{Celis:2017hod,Fuentes-Martin:2020zaz}. To linearise the system, we 
fully decouple the running of SM parameters from that of the dimension-6 Wilson 
coefficients (WCs). The SM parameters are evolved using the five-loop RGEs already 
implemented in \dsix, with input values at $\mu_0 = 10$~TeV taken from the 
\smefit--\textsc{wilson} interface~\cite{terHoeve:2025gey}. The 
WC running is then solved numerically from a fixed high scale of $10$~TeV down to 
the relevant observable scales, exploiting the interpolating function provided by 
\dsix~to evaluate the solution at all scales present in the \smefit~database. The 
running is computed retaining all SM gauge couplings, the Higgs self-coupling, and 
the top Yukawa, while all other Yukawa couplings and fermion masses are set to zero and the CKM matrix to the identity.

As a final step, the RGE matrix is rotated from the Warsaw basis used in \dsix~to 
the \smefit~basis via the dictionary implemented in the \smefit~RGE module and 
available in the WCxf repository (see Appendix~\ref{app:smefit_basis} for details). 
This yields a pre-computed two-loop RGE matrix that is fed into \smefit~via the 
corresponding runcard option, supporting both fixed and dynamical observable scales. 
Unless stated otherwise, all results presented below use a dynamical observable scale.

Consistency of the calculation requires that observables be computed in the same 
renormalization scheme used to derive the beta functions. The calculations in 
\textsc{SMEFT@NLO} \cite{Degrande:2020evl} are performed in $\overline{\mathrm{MS}}$ and adopt the same 
prescription for the projection of physical structures as~\cite{Born:2026xkr,
Fuentes-Martin:2022vvu}. Since the predictions are expressed exclusively in terms 
of physical Wilson coefficients, they implicitly define an evanescent-free scheme, or finitely-compensated evanescent scheme, 
in which evanescent operators are reduced at the appropriate loop order. Matching 
calculations performed with \sold~\cite{Guedes:2023azv,Guedes:2024vuf} and 
\mmeft~\cite{Carmona:2021xtq} follow the same conventions and 
include the one-loop reduction of evanescent operators; the two-loop reduction, 
which would enter as part of the two-loop threshold corrections, is beyond the 
scope of this work.

We stress that our setup is perturbatively incomplete in a strict sense: a fully 
consistent fixed-order two-loop analysis would additionally require two-loop 
expressions for all observables and two-loop matching conditions for the UV models 
of interest, neither of which is currently available. The present study should 
therefore be understood as a preliminary assessment, the first of its kind, 
of the phenomenological impact of two-loop RGE effects in SMEFT global fits, 
providing both a quantitative estimate of the size of corrections and a guide for 
future complete calculations.

\afterpage{
    \thispagestyle{empty}
    \begin{figure}[H]
        \centering
        \begin{adjustbox}{center}
            \includegraphics[width=1.2\linewidth]{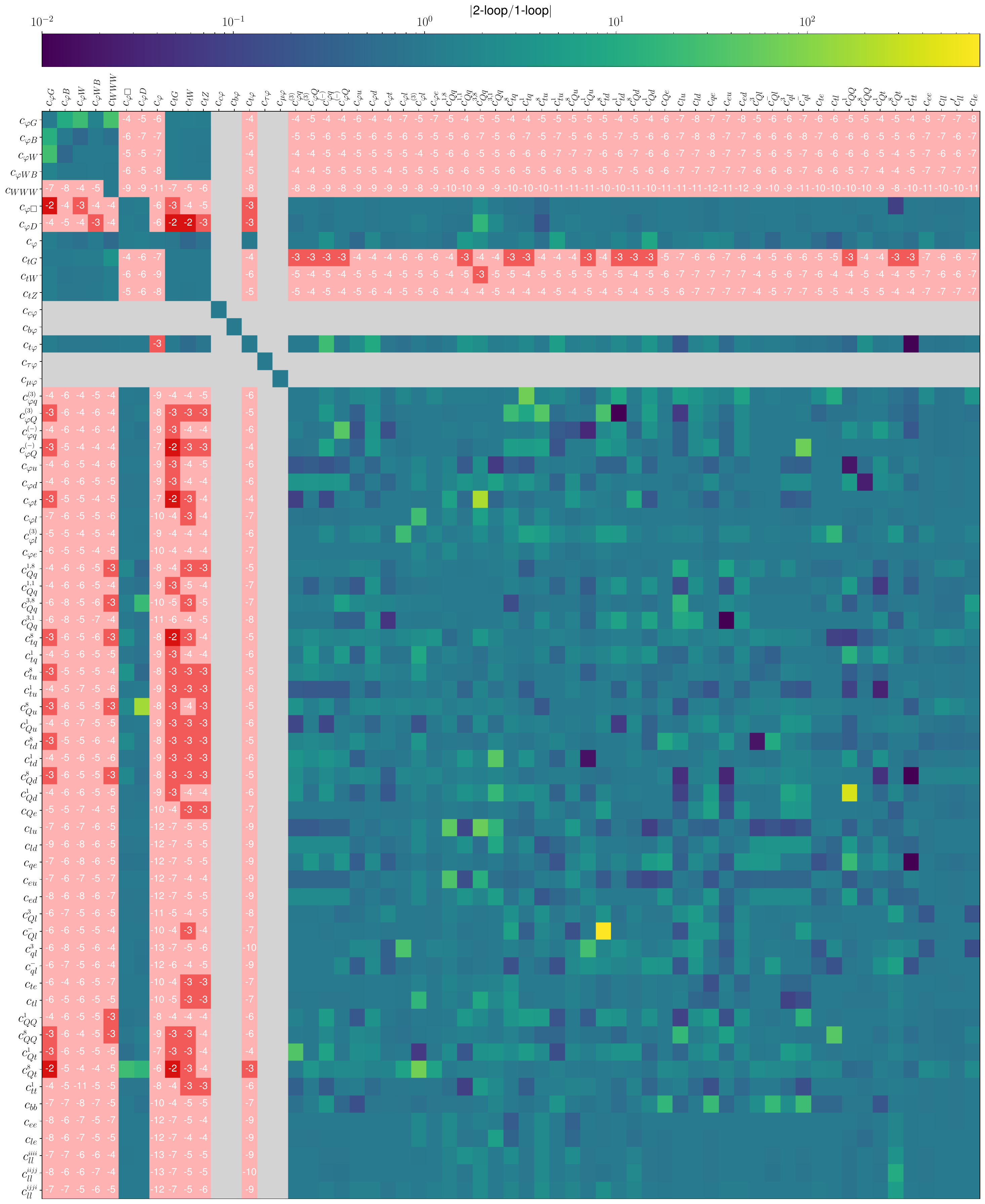}
        \end{adjustbox}
        \caption{Ratio of the two-loop to one-loop RGE evolution matrices for running 
        between $10$~TeV and $m_{Z}=91.19$~GeV, displayed as a heatmap in the 
        \smefit~basis. The $(i,\,j)$ entry represents the mixing of $c_{j}$, defined 
        at the high scale of $10$~TeV, into $c_{i}$, evaluated at the observable scale 
        $m_{Z}$. Entries shown in red indicate one-loop zeros broken at two loops, with 
        the base-10 logarithm of the newly induced mixing indicated in white; darker 
        shades correspond to larger mixings. Entries shown in gray are zeros of the 
        two-loop RGE matrix.}
        \label{fig:heatmap_twoloop_changes}
    \end{figure}
}

\subsection{Comparison with the one-loop matrix}
\label{sec:comparison_1_vs_2_loops}

To assess the structural impact of two-loop corrections, we compute the SMEFT 
evolution matrix between $10$~TeV and $m_{Z}=91.19$~GeV at both one and two loops 
and compare the results. The ratio of the two solutions in the \smefit~basis is 
displayed as a heatmap in Fig.~\ref{fig:heatmap_twoloop_changes}, where the 
$(i,\,j)$ entry represents the mixing of $c_{j}$ into $c_{i}$: that is, $c_{j}$ 
is defined at the high scale of $10$~TeV, while $c_{i}$ is evaluated at the 
observable scale $m_{Z}$. Under the flavor symmetry assumed throughout this work, operators differing only 
in lepton flavor index share identical mixing patterns and are therefore grouped 
into a single coefficient in the figure.

We begin with the zero structure of the two matrices. The two-loop RGE matrix 
retains 924 exact zeros, shown in gray in Fig.~\ref{fig:heatmap_twoloop_changes}, 
approximately one third of those present at one loop. All remaining zeros are associated with dimension-6 Yukawa operators of 
fermions other than the top, and follow directly from our assumption that the top 
Yukawa is the only non-vanishing SM Yukawa. At two loops, the mixing structure becomes 
significantly denser, with 1896 previously vanishing entries acquiring non-zero 
contributions. These newly induced mixings are highlighted in red in 
Fig.~\ref{fig:heatmap_twoloop_changes}, with the base-10 logarithm of their 
magnitude indicated in white: an entry of order $X \times 10^{-3}$ is thus labeled 
$-3$. The most phenomenologically relevant of these are discussed below.

The one-loop result that $c_{\varphi}$ does not run into other operators is lost 
at two loops, though most of the new mixings are very small; a notable exception 
is a $10^{-3}$ mixing into $c_{t\varphi}$. Similarly, $c_{WWW}$ acquires new 
mixings into two- and four-heavy four-quark operators at order $10^{-3}$. The 
operator $c_{\varphi G}$, which mixes into very few operators at one loop, develops 
order-$10^{-3}$ mixings with $c_{\varphi t}$, $c_{\varphi Q}^{(-)}$, 
$c_{\varphi Q}^{(3)}$, $c_{td}^{8}$, $c_{Qd}^{8}$, $c_{tq}^{8}$, $c_{Qu}^{8}$, 
$c_{tu}^{8}$, $c_{QQ}^{8}$, $c_{Qt}^{1}$, and order-$10^{-2}$ mixings with 
$c_{Qt}^{8}$ and $c_{\varphi\Box}$. Several further one-loop zeros are broken at 
order $10^{-3}$: $c_{\varphi Q}^{(-)}$ mixes into the top dipoles (including the 
chromomagnetic one); $c_{\varphi W}$ mixes into $c_{\varphi\Box}$; $c_{\varphi WB}$ 
mixes into $c_{\varphi D}$; and $c_{\varphi q}^{(-)}$ and $c_{tt}^{1}$ both 
acquire new mixings into $c_{tG}$.

The dipole sector, which forms a largely isolated block at one loop, is 
significantly more connected at two loops. The top dipoles acquire new mixings into 
a broad range of four-fermion operators, including four-heavy-quark, 
two-heavy-two-light-quark, and two-heavy-quark-two-lepton operators, all at order 
$10^{-3}$. No clear hierarchy is observed between the QCD and electroweak dipole 
mixings into four-quark operators, but the electroweak dipoles mix more strongly 
than the QCD dipole into two-quark-two-lepton and four-lepton operators. 
Particularly significant are the new mixings of all three dipoles into 
$c_{\varphi D}$, reaching order $10^{-2}$ for $c_{tG}$ and $c_{tW}$ and order 
$10^{-3}$ for $c_{tZ}$. The operator $c_{tG}$ additionally develops 
order-$10^{-2}$ mixings into $c_{\varphi Q}^{(-)}$ and $c_{\varphi t}$, and 
order-$10^{-3}$ mixings into their light-quark counterparts. 
Note, however, that the mixings of all dipoles and coefficients of the $c_{\varphi X}$ class are formally of higher loop order because they cannot be generated at tree level by any weakly-coupled extension.
Finally, $c_{t\varphi}$, 
which at one loop mixes only into $c_{\varphi}$, acquires mixings into virtually 
all non-Yukawa operators at two loops, though most remain very small; the 
phenomenologically most relevant are order-$10^{-3}$ mixings into $c_{Qt}^{8}$, 
$c_{\varphi\Box}$, and $c_{\varphi D}$, the latter two being tightly constrained 
by data.

We now turn to the corrections to non-vanishing one-loop mixings. A first glance 
at Fig.~\ref{fig:heatmap_twoloop_changes} might suggest that most changes are at 
the percent level or below, but this impression is misleading: $68\%$ of all 
non-vanishing one-loop mixings receive corrections of at least $5\%$ from two-loop 
diagrams, with a median correction of $27\%$. 

The most dramatic changes are the order-$75000\%$ corrections to the mixing of 
$c_{td}^{8}$ into $c_{Q\ell_{1,2,3}}^{-}$, visible as the brightest yellow entries 
in Fig.~\ref{fig:heatmap_twoloop_changes}. In the opposite direction, 27 mixings 
(counting lepton flavor multiplicities) are suppressed by two-loop effects at the 
level of $10^{-2}$, including the mixing of $c_{td}^{1}$ into $c_{\varphi Q}^{(3)}$, 
of $c_{(e/\mu/\tau)u}$ into $c_{Qq}^{3,1}$, and of $c_{tt}^{1}$ into $c_{t\varphi}$, 
$c_{q(e/\mu/\tau)}$, and $c_{Qd}^{8}$. 
Nevertheless, we emphasize that these large relative changes occur 
almost exclusively in matrix elements that were already small at one loop, with 
only $0.5\%$ of the significantly corrected elements having a one-loop value above 
$0.05$.

A deeper understanding of the origin of 
these extreme cancellations and enhancements is left for future work; for the 
purposes of this paper, the key finding is that two-loop corrections to the RGE 
matrix are structurally significant and cannot be treated as a uniform perturbation 
of the one-loop result.

\section{Effects on bottom-up SMEFT fits}
\label{sec:smeft_fits}

Having characterized the structural differences between the one- and two-loop RGE 
matrices, we now turn to their phenomenological impact in the context of global 
SMEFT fits. We perform all fits using the latest release of 
\smefit~\cite{Giani:2023gfq}, with a dataset updated to include the projections 
from the European Strategy for Particle Physics Update~\cite{Armadillo:2026xyz,deBlas:2025gyz}.

The dataset combines current LHC measurements with future collider projections. On 
the LHC side, we include measurements of single Higgs, Higgs pair, and Higgs STXS 
production, single top, $t\bar{t}$, $t\bar{t}b\bar{b}$, $t\bar{t}t\bar{t}$, 
diboson, neutral-current Drell-Yan, and $tW/tZ$ processes, together with their 
projected HL-LHC counterparts obtained following the methodology of 
\cite{Celada:2024mcf}, and dedicated HL-LHC projections for differential 
$t\bar{t}$, single Higgs, and di-Higgs production~\cite{Armadillo:2026xyz}. On 
the FCC-ee side, we include the full set of observables foreseen in the Feasibility 
Study~\cite{FCC:2025lpp} at center-of-mass energies of $91$, $161$, $240$, and 
$365$~GeV. These comprise Electroweak Precision Observables, light difermion 
production, Higgstrahlung, vector boson fusion, Higgs decays, and inclusive and 
fully differential $W^{+}W^{-}$ and $t\bar{t}$ production via Optimal Observables.
We include theoretical uncertainties on LEP and FCC-ee observables under the aggressive projections scenario~\cite{deBlas:2025gyz}.

\begin{figure}[ht]
    \centering
    \includegraphics[width=1.0\linewidth]{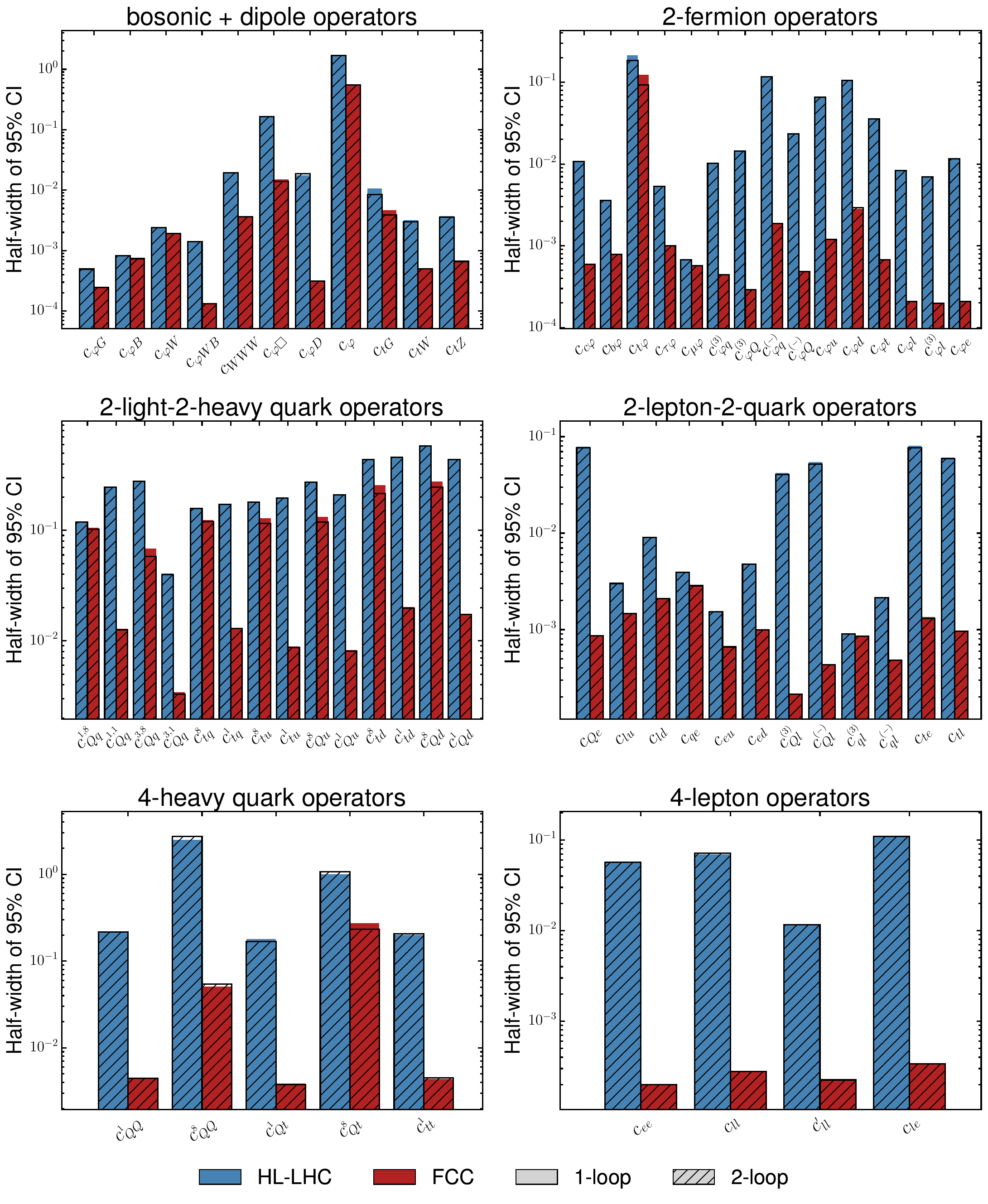}
    \caption{Half-width of the $95\%$ C.I. for dimension-6 SMEFT operators from 
    a linear individual fit, comparing results obtained with HL-LHC (blue) and 
    FCC-ee (red) projections. Solid bars correspond to one-loop RGEs; hatched bars 
    indicate the result of including the full two-loop SMEFT RGEs.}
    \label{fig:lin_ind}
\end{figure}

\begin{figure}[ht]
    \centering
    \includegraphics[width=1.0\linewidth]{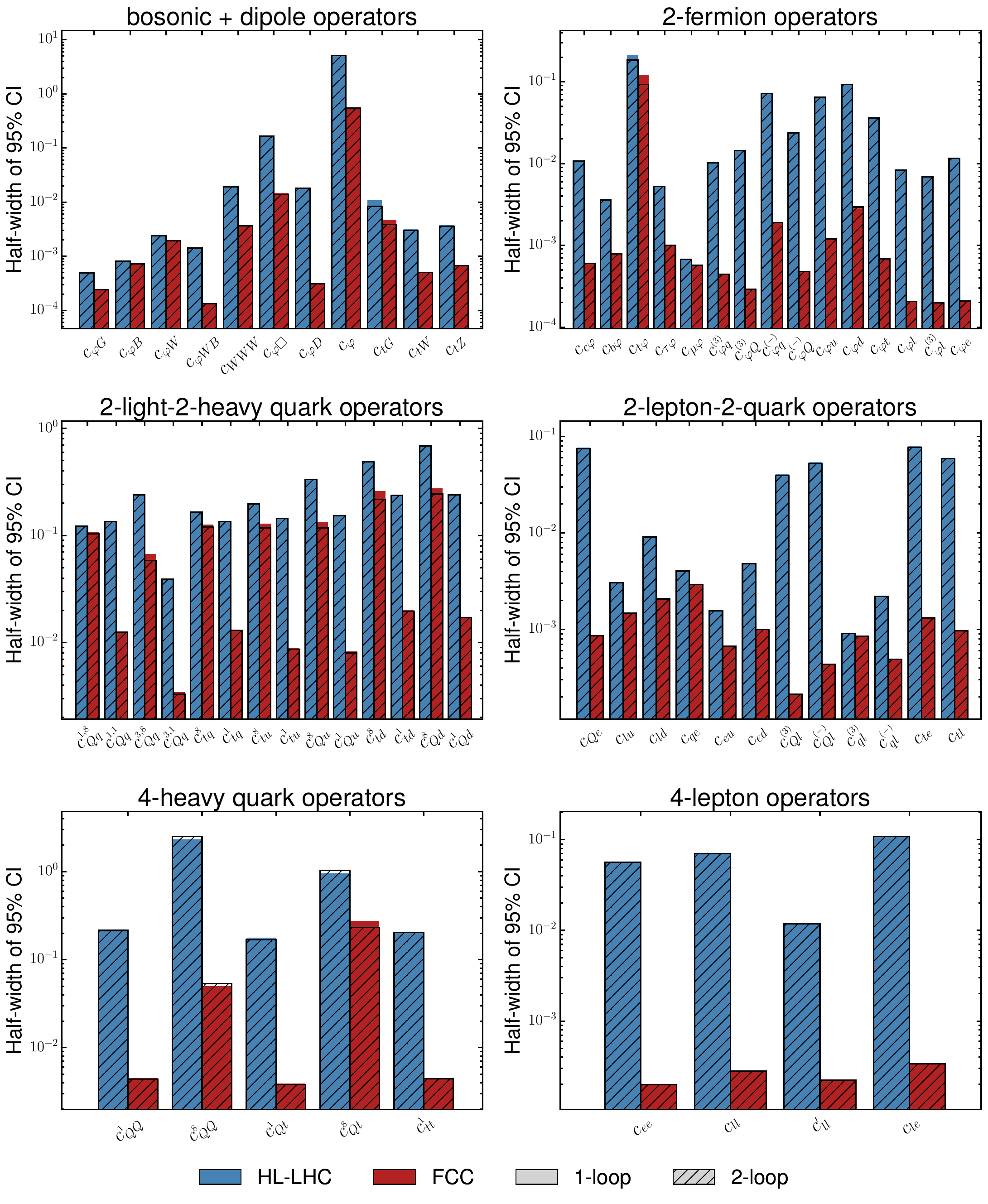}
    \caption{Half-width of the $95\%$ C.I. for dim-6 SMEFT operators from a quadratic individual fit. We show the result of using HL-LHC (blue) or FCC-ee (red) projections. The full colored bar corresponds to including RGE effects only to one-loop order, while the hatched bar indicates the result of including the full two-loop SMEFT RGEs.}
    \label{fig:quad_ind}
\end{figure}
\clearpage

The predictions are computed following the \smefit~and \textsc{SMEFT@NLO} flavour symmetry,
U$(2)_q\times$U$(2)_u\times$U$(3)_d\times($U$(1)_\ell\times$U$(1)_e)^3$, plus the symmetry-breaking charm, bottom, tau and muon dimension-6 Yukawas.
Moreover, the Drell-Yan and FCC-ee difermion production predictions have been computed with the more general U$(2)_d$ symmetry in the down-quark sector. The RGE matrix computed in the previous sections follows this more general symmetry. However, for simplicity, the fits in this section have been computed imposing a more restrictive lepton-flavor universal symmetry, U$(3)_\ell\times$U$(3)_e$, while keeping U$(2)_q\times$U$(2)_u\times$U$(3)_d$ in the quark sector.
We refer to~\cite{Armadillo:2026xyz} for further details on the setup.

\subsection{Individual fits}
\label{sec:indiv_fits}
We begin by evaluating the impact of the two-loop SMEFT RGEs on individual fits 
at HL-LHC and FCC-ee. Figure~\ref{fig:lin_ind} shows the half-width of the $95\%$ 
confidence interval (C.I.) on the dimension-6 WCs from a linear, 
$\mathcal{O}(\Lambda^{-2})$, individual fit, comparing results obtained with one- 
and two-loop RGEs.

In the bosonic and two-fermion current sectors, the two-loop RGEs have no 
appreciable impact on the sensitivity, with a few notable exceptions. 
The top-yukawa modifier $c_{t\varphi}$, whose significantly richer two-loop mixing pattern, in particular the mixing into $c_{\varphi D}$, leads 
to constraints improved by $14\%$ at HL-LHC and $25\%$ at FCC-ee, is the most 
affected. Visible improvements are observed for $c_{tG}$, 
with bounds tightening by $22\%$ and $17\%$ at HL-LHC and FCC-ee respectively, and 
for $c_{\varphi\Box}$, whose FCC-ee sensitivity improves by $7\%$.

The four-fermion sector shows a richer phenomenology. Several two-light-two-heavy 
quark operators benefit from improved FCC-ee sensitivity at two loops, driven by 
their newly induced mixing into the dipole operators. Specifically, the bounds on $c_{Qq}^{3,8}$, $c_{td}^{8}$ 
improve by $15\%$, those on $c_{Qu}^{8}$, $c_{tu}^{8}$ by $10\%$, 
and those on $c_{Qd}^{8}$ by $12\%$ at FCC-ee. Among the 
four-heavy-quark operators, $c_{Qt}^{8}$ stands out, with bounds improving 
by $15\%$ at FCC-ee, driven by its two-loop running into 
$c_{tG}$. No significant effects are observed in the two-lepton-two-quark or 
four-lepton sectors.

The inclusion of quadratic dependencies on the WCs, i.e. $\mathcal{O}(\Lambda^{-4})$ 
corrections, does not significantly alter the picture described above, as shown in 
Fig.~\ref{fig:quad_ind}. All improvements driven by two-loop RGEs persist at 
quadratic order, which is expected since in individual fits most operators remain 
in the linear regime and the quadratic corrections are subleading.

\subsection{Global fits}
\label{sec:global_fits}

As observed in previous studies~\cite{Bartocci:2024fmm,terHoeve:2025gey,deBlas:2025xhe}, the phenomenological impact of RGE running 
can differ substantially between individual and global fits, even when the latter 
involve only a handful of operators. Here, we perform a global fit with 61 
independent operators active at $10$~TeV, and find that the two-loop RGEs indeed 
leave a considerably stronger imprint than in the individual case.

\begin{figure}[h]
    \centering
    \includegraphics[width=1.0\linewidth]{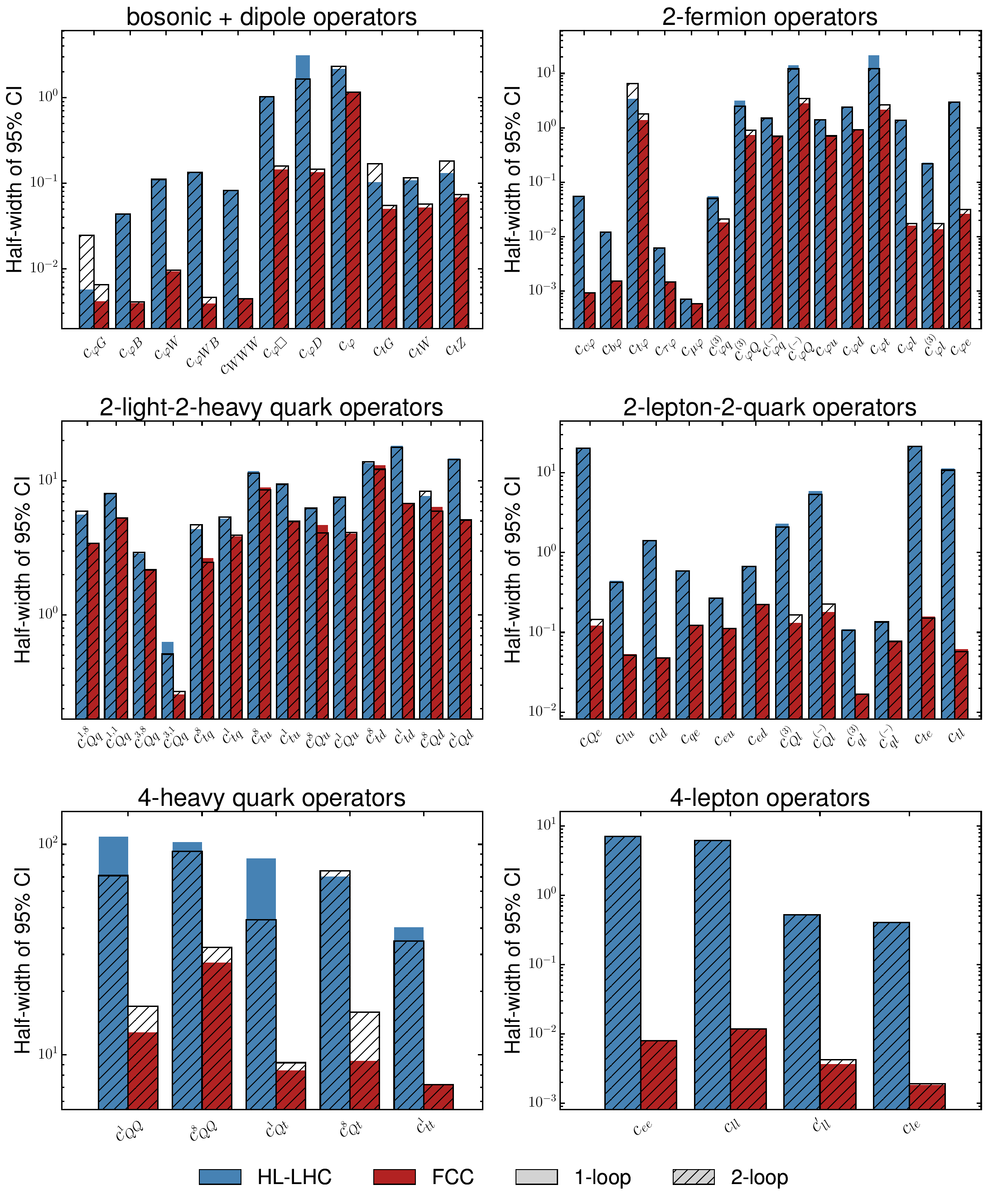}
    \caption{Half-width of the $95\%$ C.I. for dimension-6 SMEFT operators from 
    a linear global fit, comparing results obtained with HL-LHC (blue) and FCC-ee 
    (red) projections. Solid bars correspond to one-loop RGEs; hatched bars 
    indicate the result of including the full two-loop SMEFT RGEs.}
    \label{fig:lin_glob}
\end{figure}

\begin{figure}
    \centering
    \includegraphics[width=0.5\linewidth]{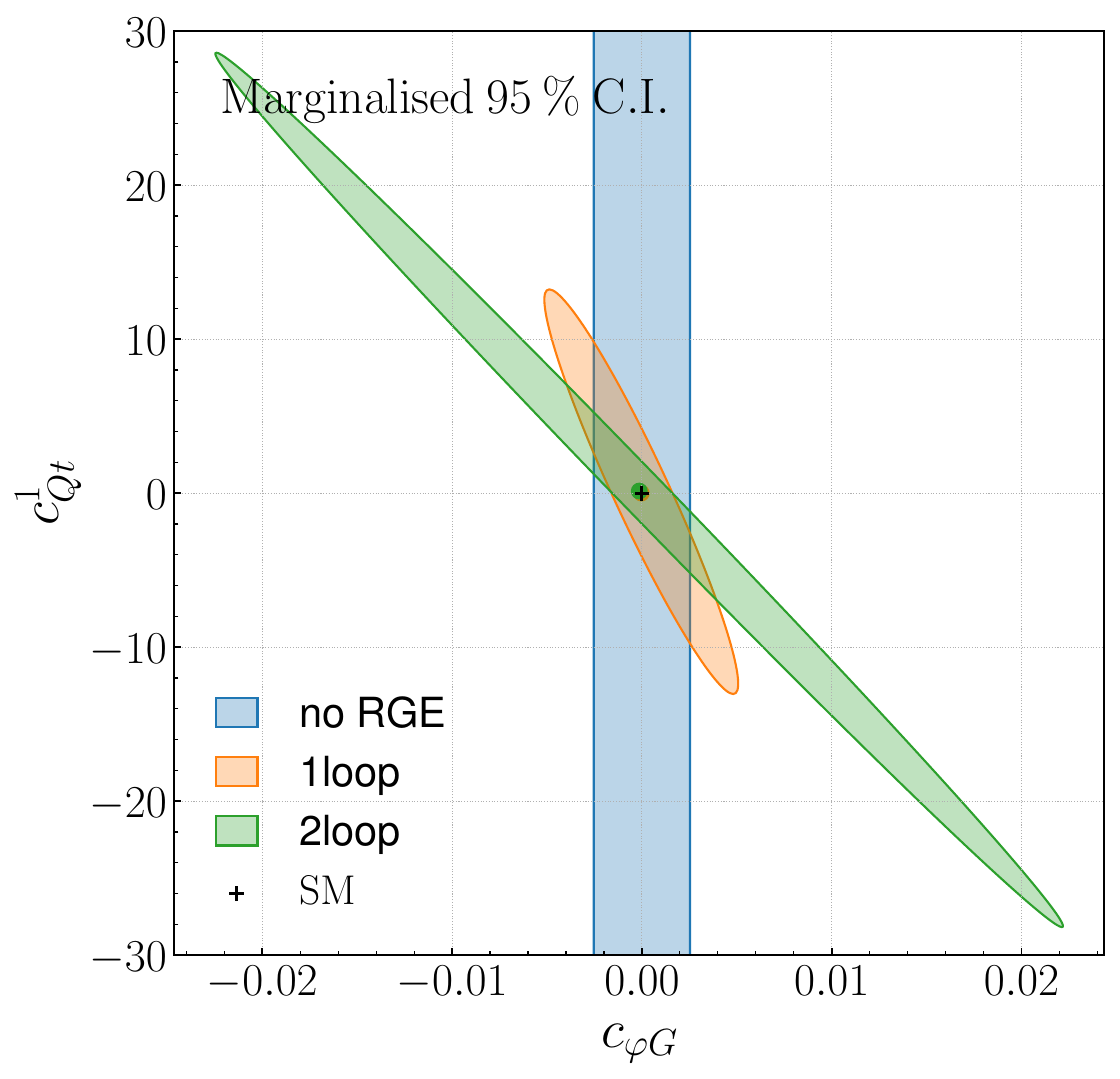}
    \caption{Two-dimensional $95\%$ credible regions in the 
    $c_{Qt}^{1}$--$c_{\varphi G}$ plane from a linear fit to Higgs signal strength 
    measurements at the HL-LHC, comparing absence of RGE, one-loop and two-loop
    RGEs.}
    \label{fig:2D-contour}
\end{figure}

Considering first the linear, $\mathcal{O}(\Lambda^{-2})$, shown in 
Fig.~\ref{fig:lin_glob}, the most striking effect concerns $c_{\varphi G}$, whose 
bounds worsen by a factor of $\sim4$ at HL-LHC and by $50\%$ at FCC-ee when switching from 
one- to two-loop RGEs. This degradation can be traced to the two-loop induced 
mixing of four-heavy-quark operators into $c_{\varphi G}$, which spreads the constraining power of Higgs measurements across a larger set of operators. 
The mechanism is illustrated in a toy fit in Fig.~\ref{fig:2D-contour}, which 
shows the $95\%$ credible region in the $c_{Qt}^{1}$--$c_{\varphi G}$ plane from 
a two-dimensional fit to Higgs signal strength measurements at the HL-LHC. At 
tree level, the data constrain $c_{\varphi G}$ directly but carry essentially no 
information on $c_{Qt}^{1}$. At one loop, a mild sensitivity to $c_{Qt}^{1}$ 
develops through RGE mixing, but it is only at two loops that a new mixing of 
$c_{Qt}^{1}$ into $c_{\varphi G}$ of order $10^{-4}$ opens up, as visible in 
Fig.~\ref{fig:heatmap_twoloop_changes}, triggering a strong correlation between 
the two operators. Since four-heavy-quark operators are poorly constrained in 
linear global fits and their coefficients can reach values of order $10$, their 
two-loop contribution to $c_{\varphi G}$ is of order $10^{-3}$, precisely the 
level at which the data become sensitive. Despite being numerically small in 
absolute terms, this mixing is sufficient to open a new correlation that dilutes 
the bound on $c_{\varphi G}$ significantly.
Notice how the mixing of four-quark operators into $c_{\varphi G}$ is a $\gamma_5$-scheme-dependent effect \cite{DiNoi:2023ygk, DiNoi:2025arz}, so the described effect would not happen in BMHV scheme.

Another notable effect in the bosonic sector concerns $c_{\varphi WB}$, whose 
FCC-ee sensitivity worsens by $18\%$ at two loops due to two- and four-fermion operators mixing into it.
A contrasting behaviour is found for $c_{\varphi D}$ and $c_{\varphi t}$, whose 
bounds tighten by $50\%$ and $43\%$ respectively at HL-LHC. Both effects 
share a common origin: the improved two-loop constraints on four-heavy-quark 
operators through their running into Higgs production observables break degeneracies 
among operators contributing to EWPOs, thereby sharpening the sensitivity to 
$c_{\varphi D}$, which enters at tree level, and to $c_{\varphi t}$, which 
contributes at one loop. This mechanism is further illustrated in 
Appendix~\ref{app:red_fit}, where we construct a reduced fit designed to isolate 
and amplify this effect.
Similarly to $c_{\varphi G}$, the sensitivity to $c_{tG}$ is degraded at both HL-LHC and FCC-ee, driven by the two-loop induced 
mixing of fermion current and four-fermion operators into it.
Among the electroweak dipoles, 
$c_{tW}$ is only slightly affected by the two-loop corrections, while $c_{tZ}$ acquires a considerably
looser bound at HL-LHC.

In the two-fermion operator sector, shown in the upper-right panel of 
Fig.~\ref{fig:lin_glob}, the most striking effect is a significant loosening of 
the bounds on $c_{t\varphi}$, in contrast to the improvement found in the 
individual fits, caused partially by the new mixing of $c_{\varphi}$ into it. 
This reversal is a characteristic feature of global 
fits, where RGE-induced correlations between operators can work in either direction 
depending on the interplay with the rest of the operator basis.

The two-light-two-heavy quark sector does not exhibit the same patterns seen in 
the individual fits; the impact of two-loop RGEs here is distributed differently, indicating that 
the bounds are largely affected by correlations with 
other operators in the global fit. The two-lepton-two-quark sector, by contrast, 
does show a visible effect: the bound on $c_{Q\ell}^{(-)}$ tightens by $10\%$ at 
HL-LHC while it worsens by $25\%$ at FCC-ee. The four-lepton 
sector remains largely unaffected by the two-loop RGEs, with the sole exception of the $c_{ll}^\prime$ coefficient bound which worsens by $16\%$ at FCC-ee.

The four-heavy-quark sector shows a much stronger two-loop impact than in the 
individual fits. In particular, we see that they improve at HL-LHC while they tend to worsen at FCC. 
The bound on $c_{QQ}^{1}$ improves by $35\%$ at HL-LHC and worsens by $33\%$ 
at FCC-ee, while its octet counterpart $c_{QQ}^{8}$ shows 
a milder effect in the both colliders. A similar pattern 
is observed for $c_{Qt}^{1}$ and $c_{Qt}^{8}$: the singlet bounds improve by $50\%$ 
and worsens by $10\%$ at HL-LHC and FCC-ee respectively, while the octet bound worsens by 
$6\%$ at HL-LHC and has a considerable degradation of $70\%$ at FCC-ee.
Finally, $c_{tt}^{1}$ is affected only at HL-LHC, where its bound 
tightens by $15\%$. Further investigation suggests that at the HL-LHC, the four-heavy operators gain additional sensitivity through their two-loop mixing into Higgs production observables. At the FCC-ee, by contrast, these operators are already more tightly constrained at one loop, and the two-loop effects introduce correlations with the two-light-two-heavy octet operators. Thus, the latter absorb part of the constraining power at two-loops and lead to a dilution of the four-heavy sector bounds.

\begin{figure}
    \centering
    \includegraphics[width=1.0\linewidth]{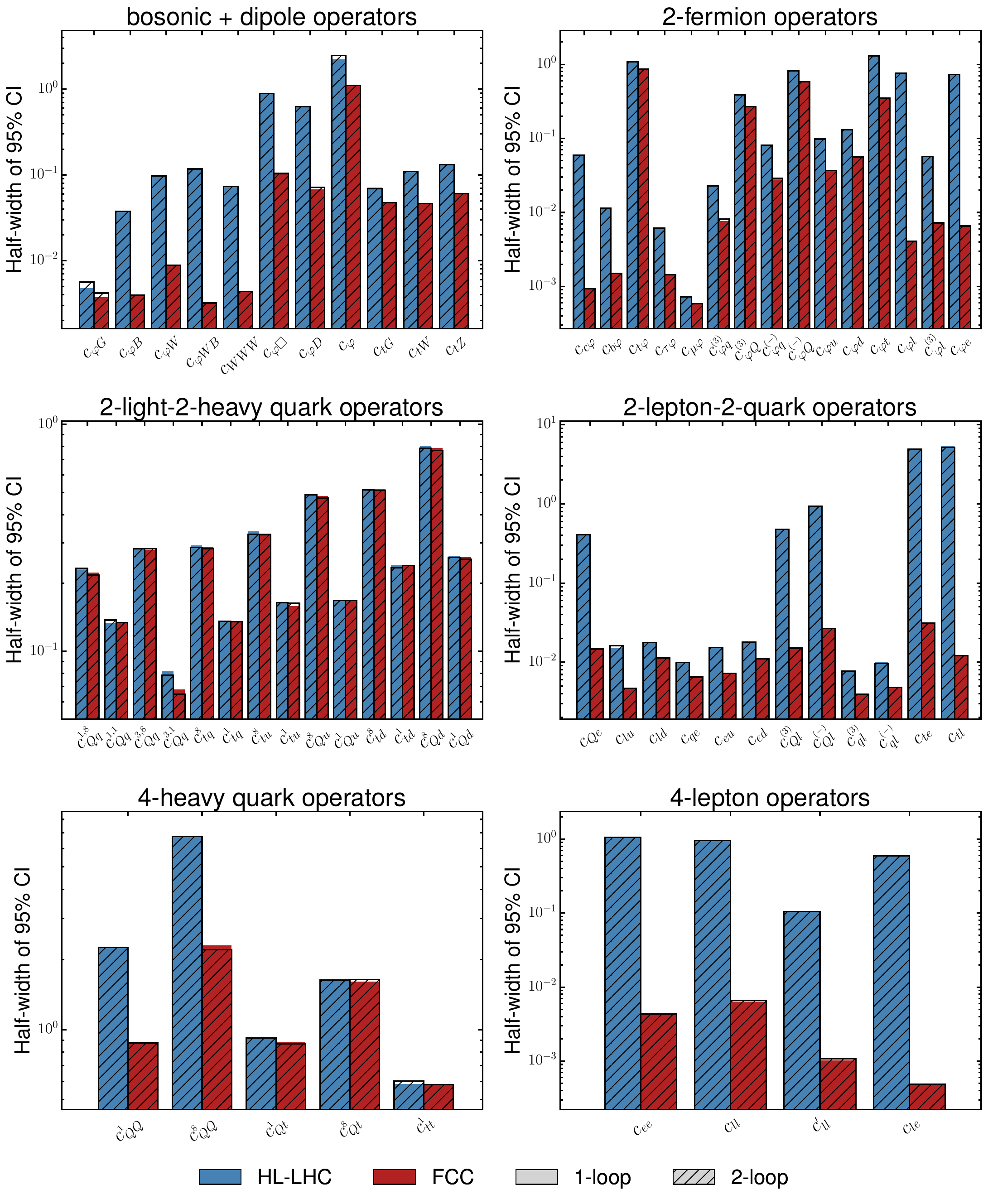}
    \caption{Half-width of the $95\%$ C.I. for dimension-6 SMEFT operators from 
    a quadratic global fit, comparing results obtained with HL-LHC (blue) and 
    FCC-ee (red) projections. Solid bars correspond to one-loop RGEs; hatched bars 
    indicate the result of including the full two-loop SMEFT RGEs.}
    \label{fig:quad_glob}
\end{figure}

When quadratic, $\mathcal{O}(\Lambda^{-4})$, corrections are included in the 
observables, many of the effects discussed above soften or disappear, as shown in 
Fig.~\ref{fig:quad_glob}. This mirrors the behaviour observed with one-loop RGEs: 
the quadratic terms break correlations among WCs, reducing the impact of 
RGE-induced mixing~\cite{Bartocci:2024fmm,terHoeve:2025gey,deBlas:2025xhe}. The 
two-loop effects that survive, albeit reduced in magnitude, include the worsening 
of the bounds on $c_{\varphi G}$ at HL-LHC and FCC-ee.

\section{Two-loop running effects in the Granada models}
\label{sec:uv_fits}

Having assessed the impact of two-loop RGEs on bottom-up SMEFT fits, we now adopt 
a top-down perspective and study their effects within concrete UV models. We focus 
on the scalar and fermionic extensions of the SM catalogued in the Granada 
dictionary~\cite{deBlas:2017xtg}. For two 
selected models of particular interest, we additionally perform a dedicated 
analysis to disentangle the effects of one-loop matching from those of two-loop 
running.
We consider the heavy scalar doublet due to a noticeable two-loop effect in the bound of its couplings, along with the $U + Q_1$ vector-like fermion pair, motivated by a previous study \cite{DiNoi:2025tka} indicating that two-loop RGEs may be relevant for this model.
The section is organized as follows: we begin with a systematic survey 
of all one-particle models in Section~\ref{sec:Granada_scalar_and_fermions}, and 
then present the dedicated model studies in 
Sections~\ref{sec:2hdm} and~\ref{sec:u_q1_mod}.

\subsection{One-particle models}
\label{sec:Granada_scalar_and_fermions}

We study the impact of two-loop SMEFT RGEs on all scalar and fermionic 
single-field extensions of the SM in the Granada dictionary~\cite{deBlas:2017xtg}. 
Vector extensions are not considered, as there is currently no systematic method 
to include loop effects that depend sensitively on the specific UV 
realization. For consistency with the flavor assumptions underlying the 
\smefit~predictions, we activate only gauge, Higgs, and third-generation fermion 
couplings, excluding $d_R$. Going beyond previous works, we also include UV 
couplings that enter exclusively through one-loop matching contributions and have 
not, to the best of our knowledge, been constrained before. Further details on the 
models and their couplings to SM fields are provided in 
App.~\ref{app:UV_details}.

All heavy fields are fixed at a mass of $M=10$~TeV, and the RGE running is 
performed from $M$ down to each observable scale at either one or two loops. For 
UV couplings entering at tree level, or to which sensitivity is found in at least 
one scenario, we adopt a flat prior on $[-100,\,100]$. All remaining couplings, 
which are practically unconstrained by the data and therefore act as nuisance 
parameters over which we marginalize, are assigned a flat prior on $[-15,\,15]$, 
motivated by the naive perturbativity bound $|g_\mathrm{UV}| < 4\pi$.

\begin{figure}[H]
    \centering
    \includegraphics[width=0.925\linewidth]{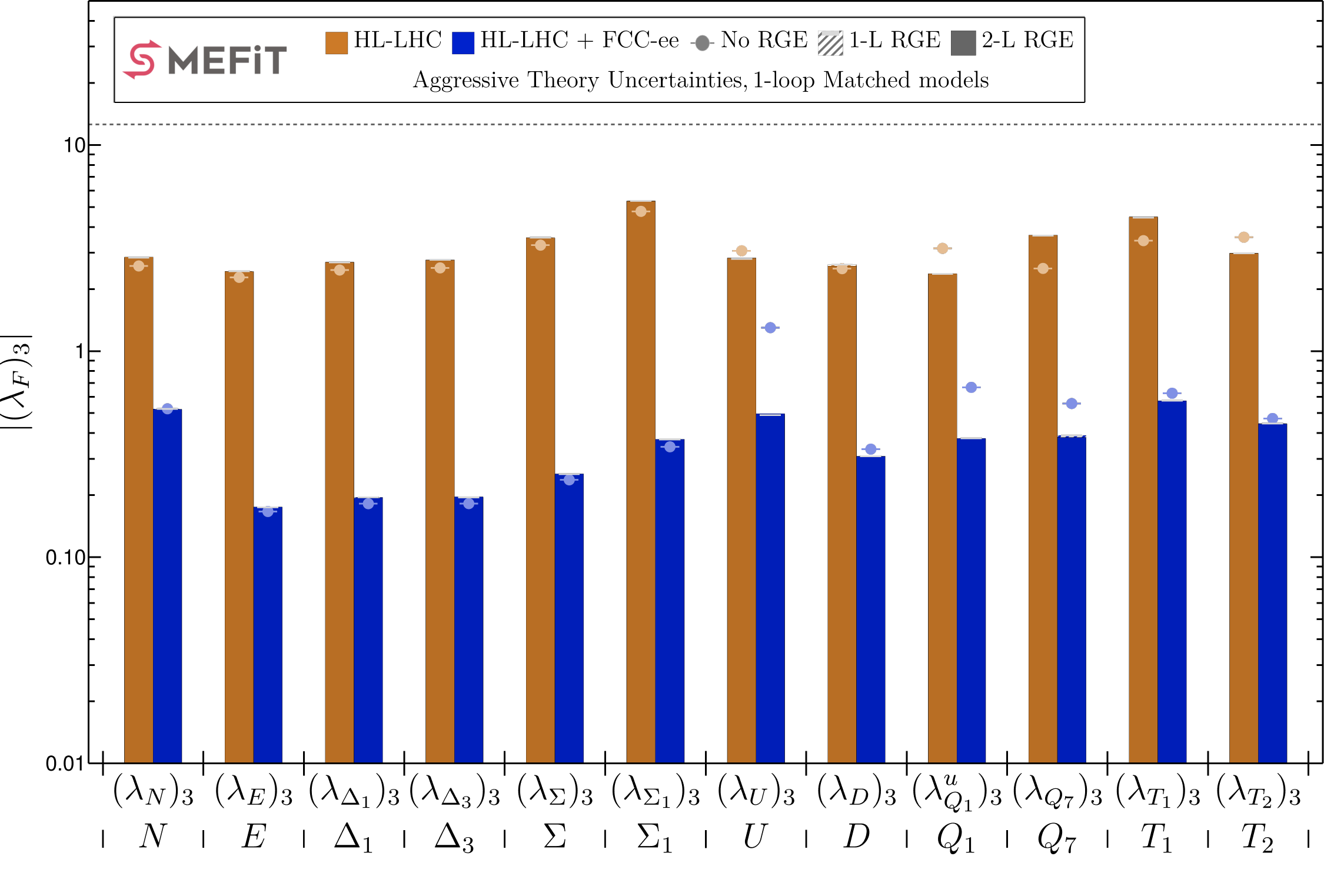}\vfill
    \includegraphics[width=0.925\linewidth]{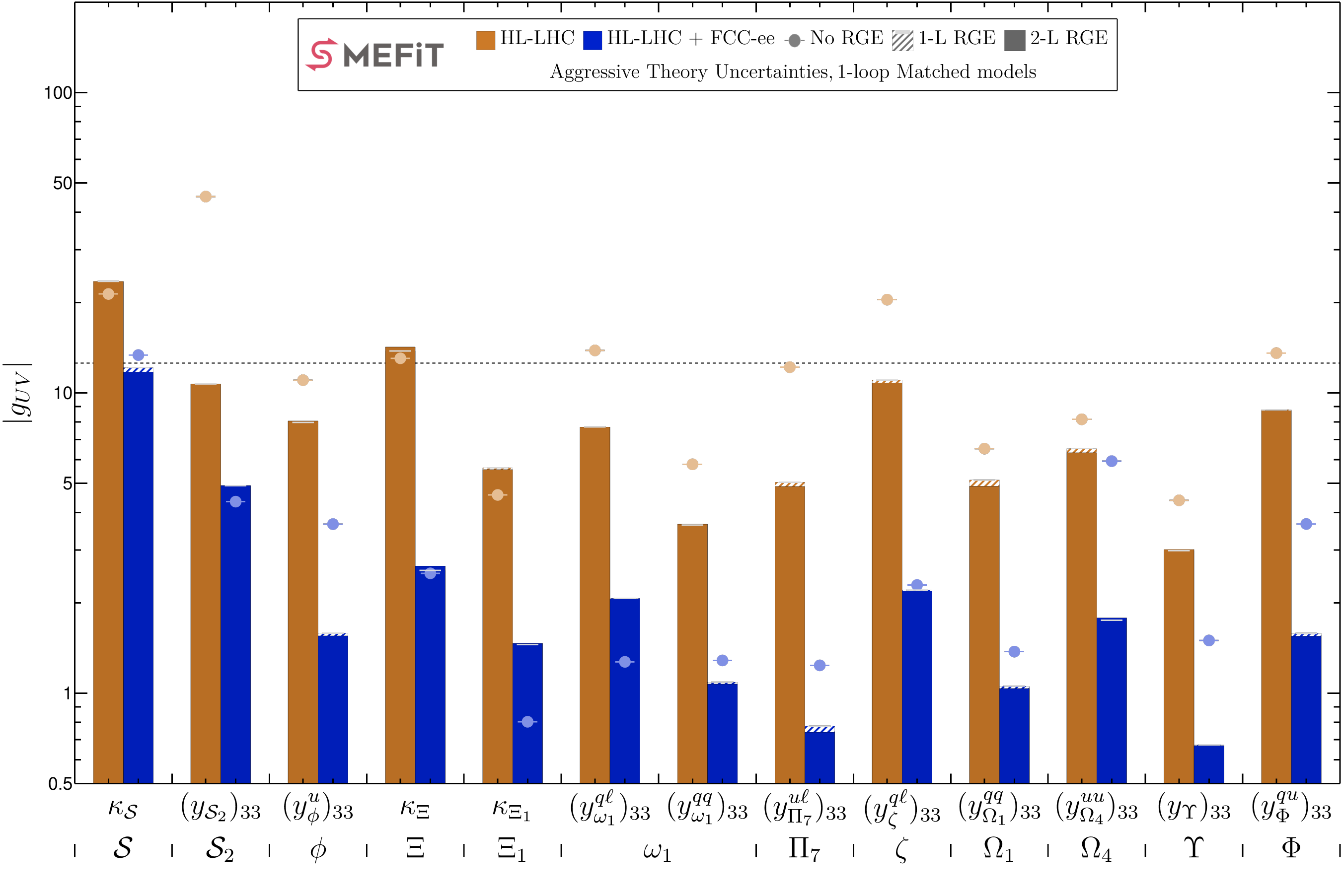}
    \caption{
    Projected $95\%$ C.I. length on the couplings of cubic interactions between heavy vector-like fermions (upper panel) or scalars (lower panel) and the SM particles. We consider one heavy particle but all its couplings at the same time (see main text) and perform the matching to the SMEFT at one loop. 
    For the case of the scalars, the bounds here include a marginalization over the quartic couplings shown in Fig.~\ref{fig:Granada_1loop_models_quartic}.
    We show the projected sensitivity at HL-LHC (dark orange) and FCC-ee (blue), assuming aggressive theory uncertainty projections, when considering no RGE effects (horizontal line with dot), one-loop SMEFT RGE (dashed bars topped with gray horizontal line) or two-loop SMEFT RGE effects (full colored bar). 
    }
    \label{fig:Granada_1loop_models_cubic}
\end{figure}

\begin{figure}[H]
    \centering
    \includegraphics[width=0.925\linewidth]{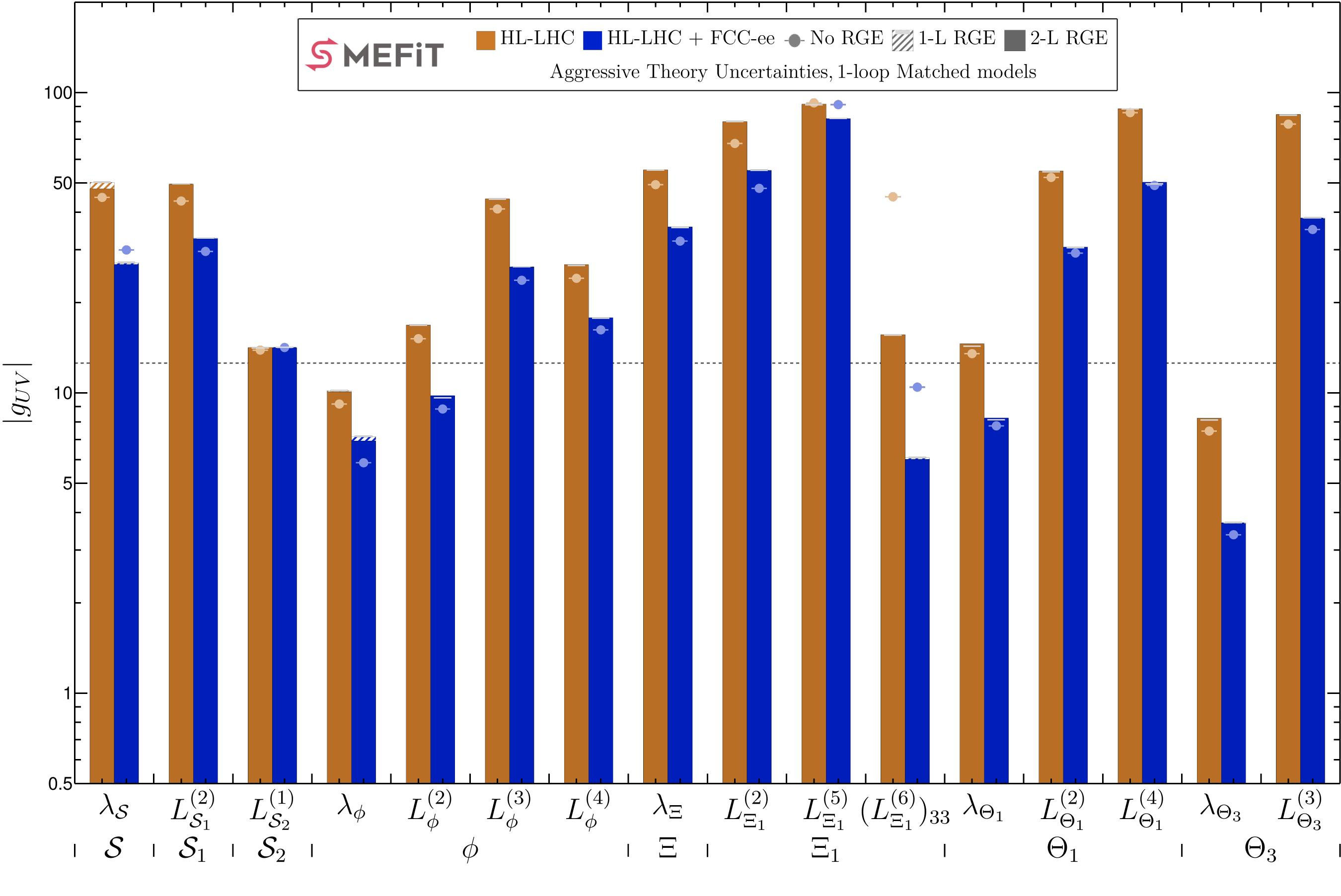}\vfill
    \includegraphics[width=0.925\linewidth]{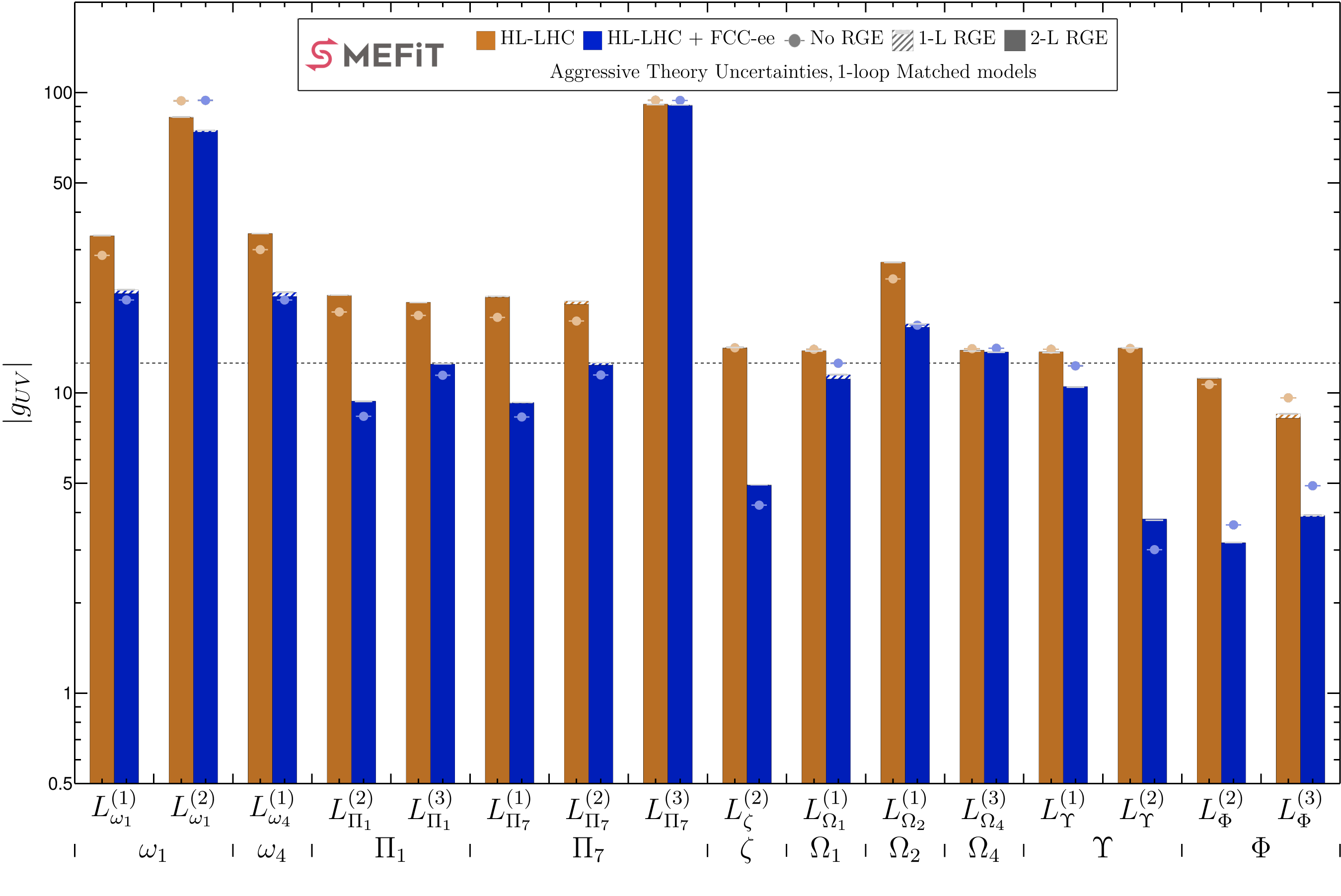}
    \caption{Projected $95\%$ C.I. length on the couplings of quartic interactions between heavy colorless (upper panel) or colored (lower panel) scalars and the SM particles. We consider one heavy particle but all its couplings at the same time (see main text) and perform the matching to the SMEFT at one loop. 
    For the case of the scalars, the bounds here include a marginalization over the cubic couplings shown in Fig.~\ref{fig:Granada_1loop_models_cubic}.
    We show the projected sensitivity at HL-LHC (dark orange) and FCC-ee (blue), assuming aggressive theory uncertainty projections, when considering no RGE effects (horizontal line with dot), one-loop SMEFT RGE (dashed bars topped with gray horizontal line) or two-loop SMEFT RGE effects (full colored bar).}
    \label{fig:Granada_1loop_models_quartic}
\end{figure}

The $95\%$ C.I. lengths on the UV couplings are shown in 
Figs.~\ref{fig:Granada_1loop_models_cubic} and~\ref{fig:Granada_1loop_models_quartic} 
for HL-LHC and FCC-ee, comparing the three RGE settings: no running, one-loop, 
and two-loop. As in the previous section, we adopt aggressive projections for the 
theoretical uncertainties throughout.

For vector-like fermions, shown in the upper panel of 
Fig.~\ref{fig:Granada_1loop_models_cubic}, the two-loop RGEs produce no 
appreciable improvement over the one-loop results. The inclusion of one-loop RGEs 
already yields significant improvements at FCC-ee, particularly for models 
coupled to the top quark, in line with previous findings~\cite{Allwicher:2024sso,
terHoeve:2025gey}. The only model where two-loop effects are marginally 
visible is $Q_7$, with an improvement of $\sim2\%$. For several models, 
particularly the colourless ones, RGE effects actually worsen the bounds slightly, 
as the self-running of tree-level-generated operators dominates over any 
RGE-induced mixing into better-constrained directions.

The results for scalar extensions are shown in the lower panel of 
Fig.~\ref{fig:Granada_1loop_models_cubic} and in 
Fig.~\ref{fig:Granada_1loop_models_quartic}. The former displays bounds on cubic 
couplings, with dimensionful couplings given in units of TeV, while the latter 
shows bounds on quartic couplings, including those entering only via one-loop 
matching, generically denoted $L_{\psi}^{(n)}$, split by the $\mathrm{SU}(3)_c$ 
charge of the scalar.

For cubic couplings, two-loop RGEs improve the sensitivity by $2$--$5\%$ in 
several cases. At HL-LHC, improvements are observed for 
$(y_{\Pi_7}^{u\ell})_{33}$, 
$\left(y_\zeta^{q\ell}\right)_{33}$, 
$\left(y_{\Omega_1}^{qq}\right)_{33}$, and
$\left(y_{\Omega_4}^{uu}\right)_{33}$. 
At FCC-ee, a two-loop improvement of $2$--$5\%$ is observed for 
$\kappa_\mathcal{S}$,
$\left(y_\phi^u\right)_{33}$,
$(y_{\Pi_7}^{u\ell})_{33}$, 
and 
$\left(y_{\Phi}^{qu}\right)_{33}$.
In some cases the two-loop RGEs can instead worsen the bound by a few percent, 
as seen for $\kappa_{\Xi}$ at HL-LHC and FCC-ee. The improvement for 
$\left(y_\phi^u\right)_{33}$ can be understood from the two-loop induced mixing 
of $c_{t\varphi}$ into tightly constrained bosonic operators, in particular 
$c_{\varphi\Box}$ and $c_{\varphi D}$. 
The remaining improvements are driven by 
the two-loop effects on four-heavy and 2-heavy-quark-2-lepton operators.

The impact of two-loop RGEs on quartic couplings, shown in 
Fig.~\ref{fig:Granada_1loop_models_quartic}, is generally smaller than for cubic 
ones. At HL-LHC, the bounds on $\lambda_\mathcal{S}$, $L_{\Pi_7}^{(2)}$ and $L_{\Phi}^{(3)}$ show a $2$--$5\%$ improvement. 
At FCC-ee, improvements of 
the same magnitude are found for $\lambda_\mathcal{S}$, $\lambda_\phi$, 
$L_{\omega_1}^{(1)}$, 
$L_{\omega_4}^{(1)}$, 
$L_{\Pi_7}^{(2)}$,
$L_{\Omega_1}^{(1)}$, and
$L_{\Omega_2}^{(1)}$. These improvements are mainly driven by contributions to $c_{\varphi\Box}$ and/or $c_{t\varphi}$.

Besides the two-loop-induced effects, we found remarkable sensitivity to some of the couplings that enter only at one-loop level, which had not been studied before. Among the colorless models, we find strong sensitivity to the couplings $L_{\phi}^{(2)}$ and $(L_{\Xi_1}^{(6)})_{33}$ at FCC-ee. The sensitivity to $L_{\phi}^{(2)}$ is driven by its contributions to $c_\varphi$ and $c_{t\varphi}$, while $(L_{\Xi_1}^{(6)})_{33}$ enters in the WC of any operator with a third-generation left-handed lepton current, including 2-quark-2-lepton and 4-lepton operators.
In the case of colored models, the FCC-ee bounds on $L_{\Pi_1}^{(2)}$, $L_{\Pi_7}^{(1)}$, $L_{\zeta}^{(2)}$, $L_{\Upsilon}^{(2)}$, $L_{\Phi}^{(2)}$ and $L_{\Phi}^{(3)}$ are $\lesssim10$, with the HL-LHC bounds for the last two also $\sim10$. The sensitivity to this latter group is driven by their contributions to bosonic WCs such as $c_{\varphi D}$ (all of them), $c_{\varphi\Box}$ (all except $L_{\Pi_1}^{(2)}$), $c_{\varphi WB}$ (all except $L_{\Pi_1}^{(2)}$ and $L_{\Pi_7}^{(1)}$), $c_{\varphi G}$ and $c_{\varphi W}$ (only $L_{\Phi}^{(2),(3)}$).

We note that for the rest of bounds on quartic couplings, most of them exceed the naive perturbativity limit of $4\pi$ 
and therefore do not yet offer physically meaningful constraints. Nevertheless, 
we include them here since two avenues exist to bring them into the perturbative 
regime: reducing the experimental or theoretical uncertainties, or lowering the 
heavy particle mass below the $10$~TeV reference value used here. Two notable 
exceptions deserve mention: without RGE effects, neither HL-LHC nor FCC-ee can 
constrain $L_{\omega_1}^{(2)}$ or $L_{\Pi_7}^{(3)}$; we display the bounds 
corresponding to $95\%$ of the prior volume in these cases to illustrate how 
RGE effects generate sensitivity where none existed before.

\subsection{2HDM}
\label{sec:2hdm}

We complement the systematic survey of the previous section with a deeper analysis 
of two selected models of particular interest, beginning with the heavy scalar 
doublet $\phi\sim(1,2)_{1/2}$ from the Granada dictionary, which corresponds to 
a Two-Higgs Doublet Model (2HDM) in the alignment and decoupling limit. As before, 
we restrict the active couplings to the top Yukawa $\left(y_{\phi}^{u}\right)_{33}$, 
the linear coupling to the SM Higgs $\lambda_{\phi}$, and the quartic scalar 
couplings that contribute via finite one-loop matching contributions, described by 
the Lagrangian
\begin{align}
    \lag_{\text{UV}}\supset &\left( \sqrt{2} \,L_{\phi}^{(2)}   (\phi^\dagger 
    \varphi)^{2}+\text{h.c.}\right) + L_{\phi}^{(3)} (\epsilon_{ab}\phi^{a} 
    \varphi^b)\, (\epsilon_{ab}\phi^{a} \varphi^b)^{*} \nonumber\\
    &+ L_{\phi}^{(4)}
    \frac{(\phi^\dagger\phi)(\varphi^\dagger\varphi)+(\phi^\dagger\varphi)
    (\varphi^\dagger\phi)}{\sqrt{2}},
\end{align}
where the normalization follows the convention of the \sold~package~\cite{
Guedes:2023azv,Guedes:2024vuf}. To isolate the respective roles of finite and 
divergent one-loop corrections, we perform the analysis at both tree-level and 
one-loop matching, in each case fixing the heavy mass to $M_{\phi}=10$~TeV. When 
one-loop matching is used, we marginalize over $L_{\phi}^{(2)}$, $L_{\phi}^{(3)}$, 
and $L_{\phi}^{(4)}$.

\begin{figure}[t!]
    \centering
    \includegraphics[width=0.49\linewidth]{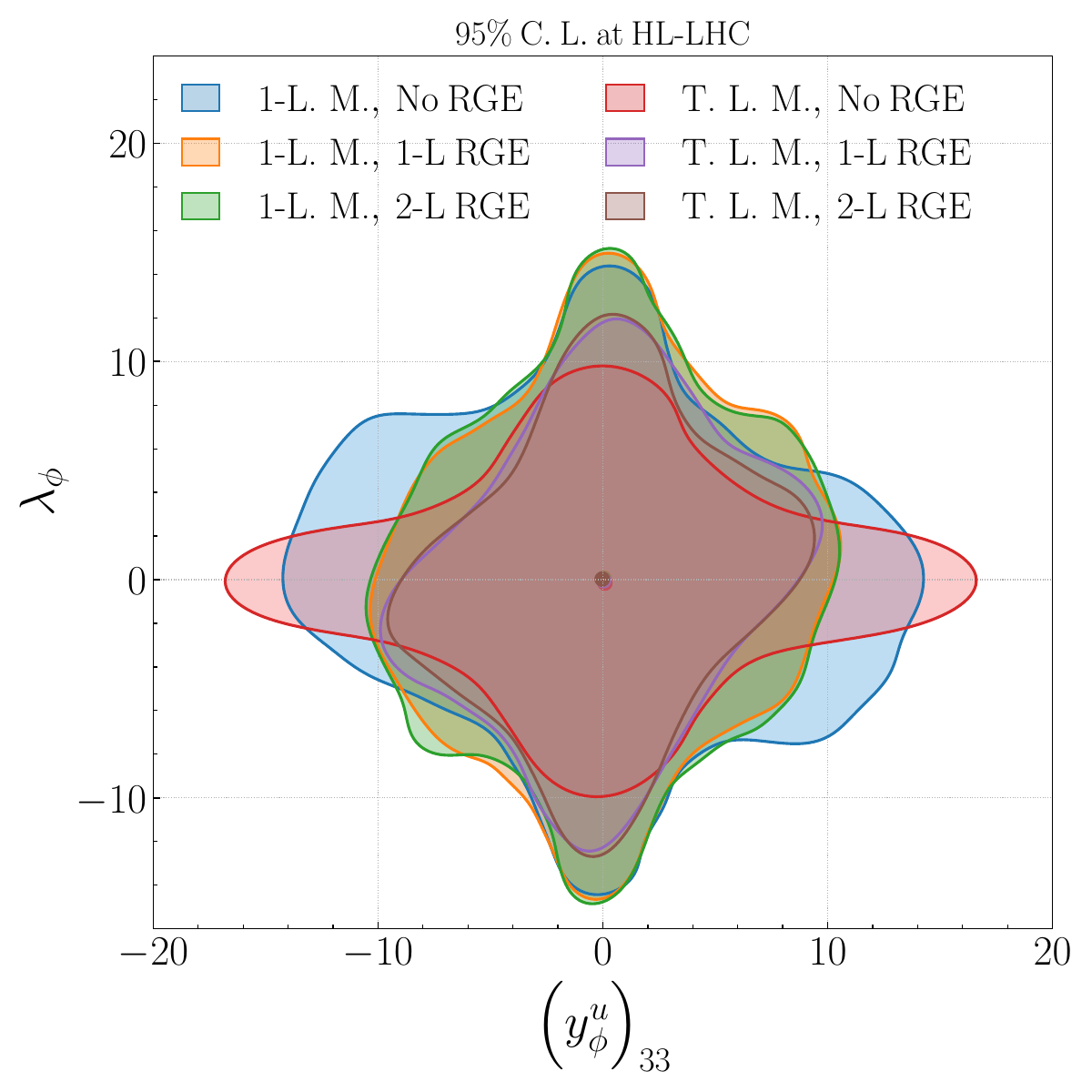}\hfill
    \includegraphics[width=0.49\linewidth]{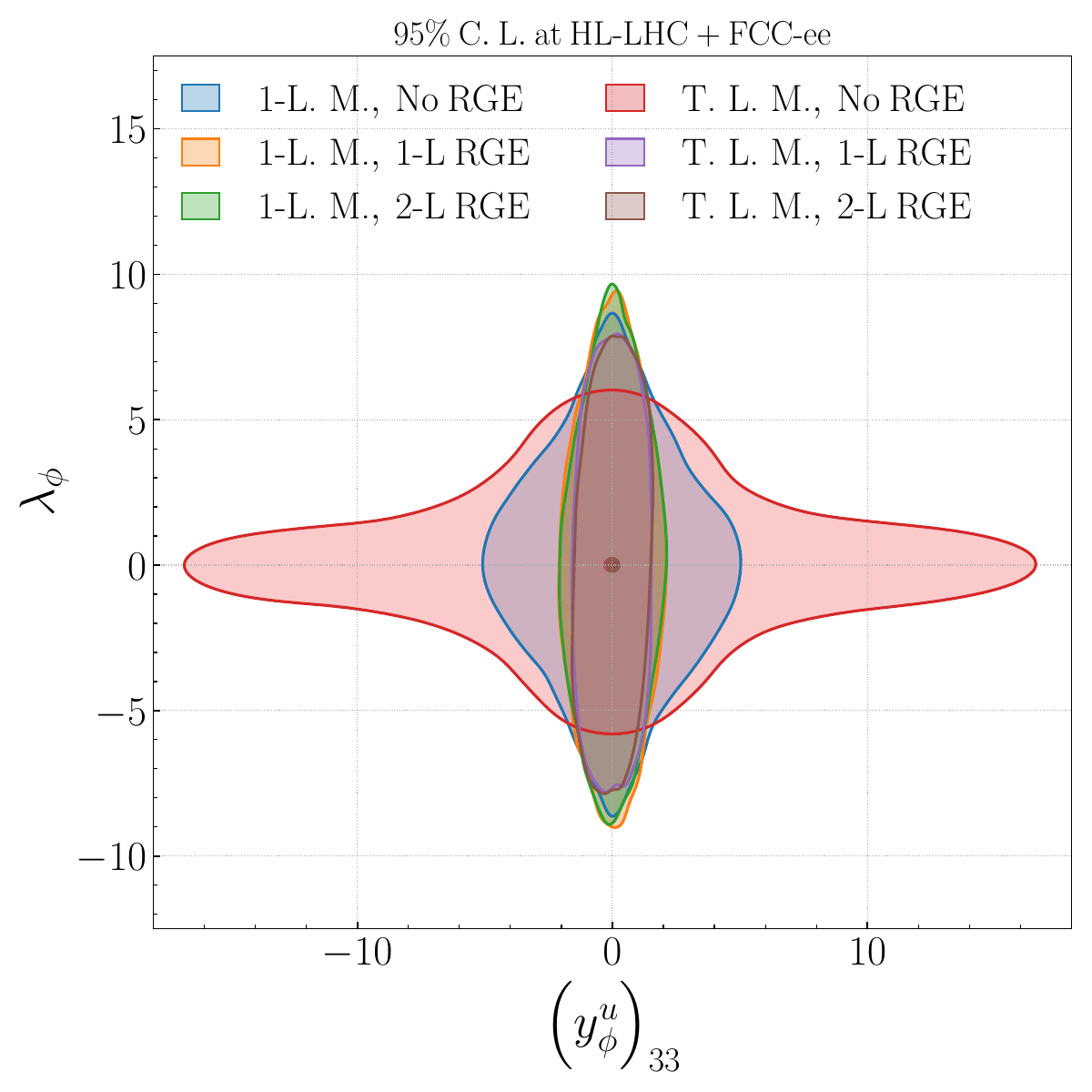}
    \caption{ Two-dimensional 
    $95\%$ 
    credible region for the UV couplings $(y_{\phi}^{u})_{33}$ and $\lambda_\phi$ for the heavy scalar doublet, $\phi$, extension of the SM at HL-LHC (left panel) and HL-LHC+FCC-ee (right panel).
    We fix the mass of $\phi$ at $M_{\phi}=10$~TeV and show the results of using tree-level (T. L. M.) and one-loop matching results (1-L. M.), and in the latter we marginalize over the UV couplings $L_{\phi}^{(2)}$, $L_{\phi}^{(3)}$, $L_{\phi}^{(4)}$.
    For each matching level, we show the bounds with none, one-loop and two-loop SMEFT RGE effects.
    We consider the aggressive theory uncertainty scenario from ESPPU26 and quadratic WC contributions.
    }
    \label{fig:2D_bounds_Varphi_model}
\end{figure}

Figure~\ref{fig:2D_bounds_Varphi_model} shows the $95\%$ credible regions in the 
$\left(y_{\phi}^{u}\right)_{33}$--$\lambda_{\phi}$ plane at HL-LHC and FCC-ee, 
for all combinations of matching level and RGE setting, under the aggressive 
theory uncertainty scenario.

At HL-LHC, the inclusion of RGE effects significantly improves the sensitivity to 
$\left(y_{\phi}^{u}\right)_{33}$, bringing the $95\%$ C.I. within the naive 
perturbativity limit, while it reduces the sensitivity to $\lambda_\phi$ due to 
the self-running of $\mathcal{O}_\varphi$. One-loop matching reproduces only part 
of the first effect, but plays a more important role in removing a spurious 
correlation along the direction $\lambda_\phi \simeq \frac{1}{4}(y_\phi^u)_{33}$ 
that appears when RGE effects are added to tree-level matching results. The 
additional precision brought by two-loop RGEs yields a modest but visible 
improvement along the directions $|\lambda_\phi| \simeq |(y_\phi^u)_{33}|$ at 
tree-level matching, and in all directions except positive $\lambda_\phi$ when 
combined with one-loop matching.

At FCC-ee, the importance of RGE effects is even more pronounced: the running of 
$\mathcal{O}_{t\varphi}$ into $Z$-pole observables generates an extreme sensitivity 
to $\left(y_\phi^u\right)_{33}$ that is entirely absent without running. In this 
case, the inclusion of one-loop matching generally reduces the overall sensitivity, 
as the three additional unconstrained directions in UV parameter space introduced 
by the quartic couplings must be marginalized over.
The combination of one-loop matching with two-loop RGE marginally improves the sensitivity to $\lambda_\phi$ at FCC-ee.

\subsection{$U+Q_1$}
\label{sec:u_q1_mod}

We conclude our model study with an analysis of the combined extension containing 
the heavy vector-like fermions $U\sim(3,1)_{2/3}$ and $Q_{1}\sim(3,2)_{1/6}$,
motivated by a previous study~\cite{DiNoi:2025tka}, where two-loop RGEs were found to improve the sensitivity, although only when combined with tree-level matching.
We consider couplings to the top quark only, fix $M_{Q_1}=M_{U}=10$~TeV, and include the full HL-LHC and FCC-ee 
datasets under the aggressive theory uncertainty scenario. 
As before, we perform the analysis at both tree-level and one-loop matching, and 
compare the results obtained with no running, one-loop, and two-loop SMEFT RGEs.

\begin{figure}[ht!]
    \centering
    \includegraphics[width=0.49\linewidth]{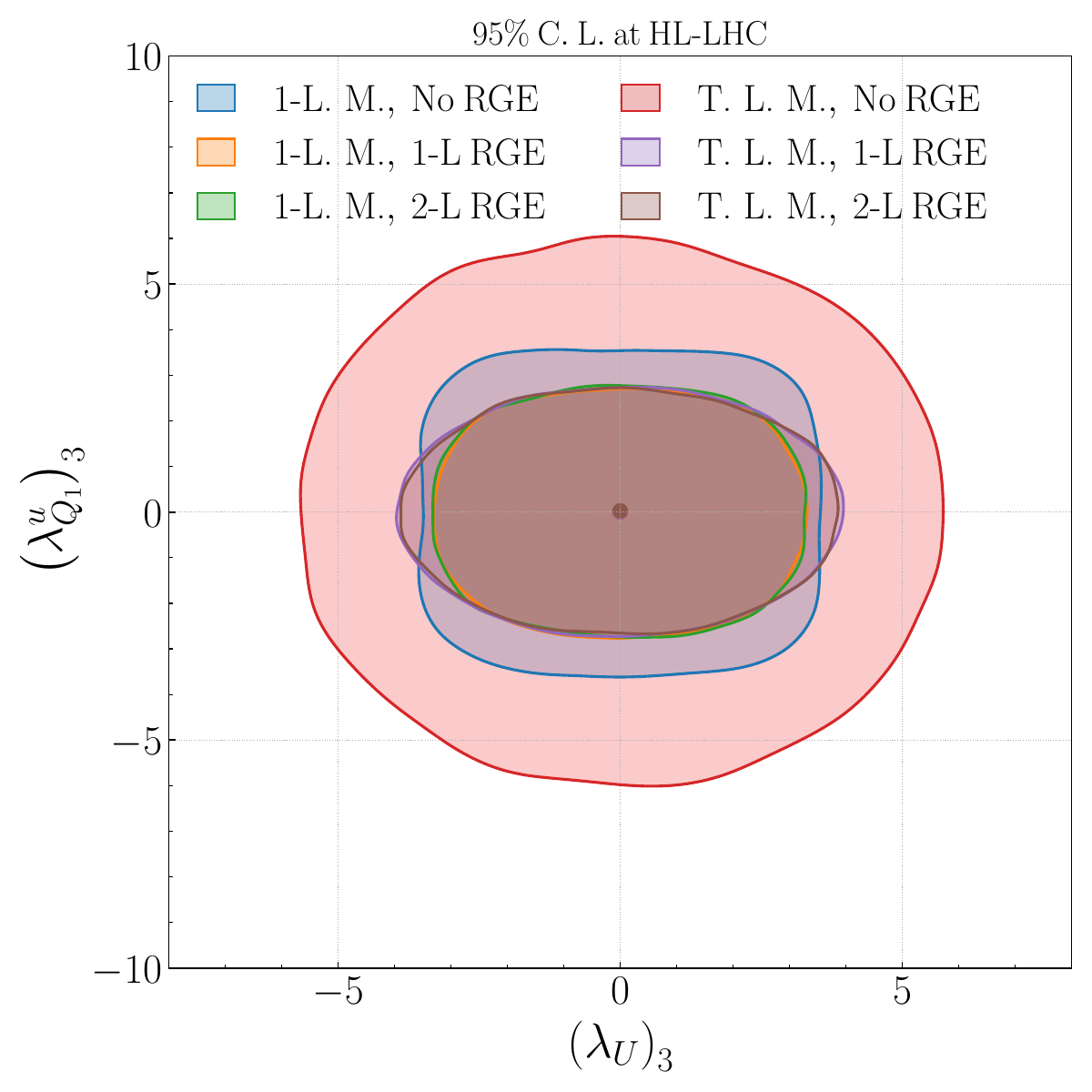}\hfill
    \includegraphics[width=0.49\linewidth]{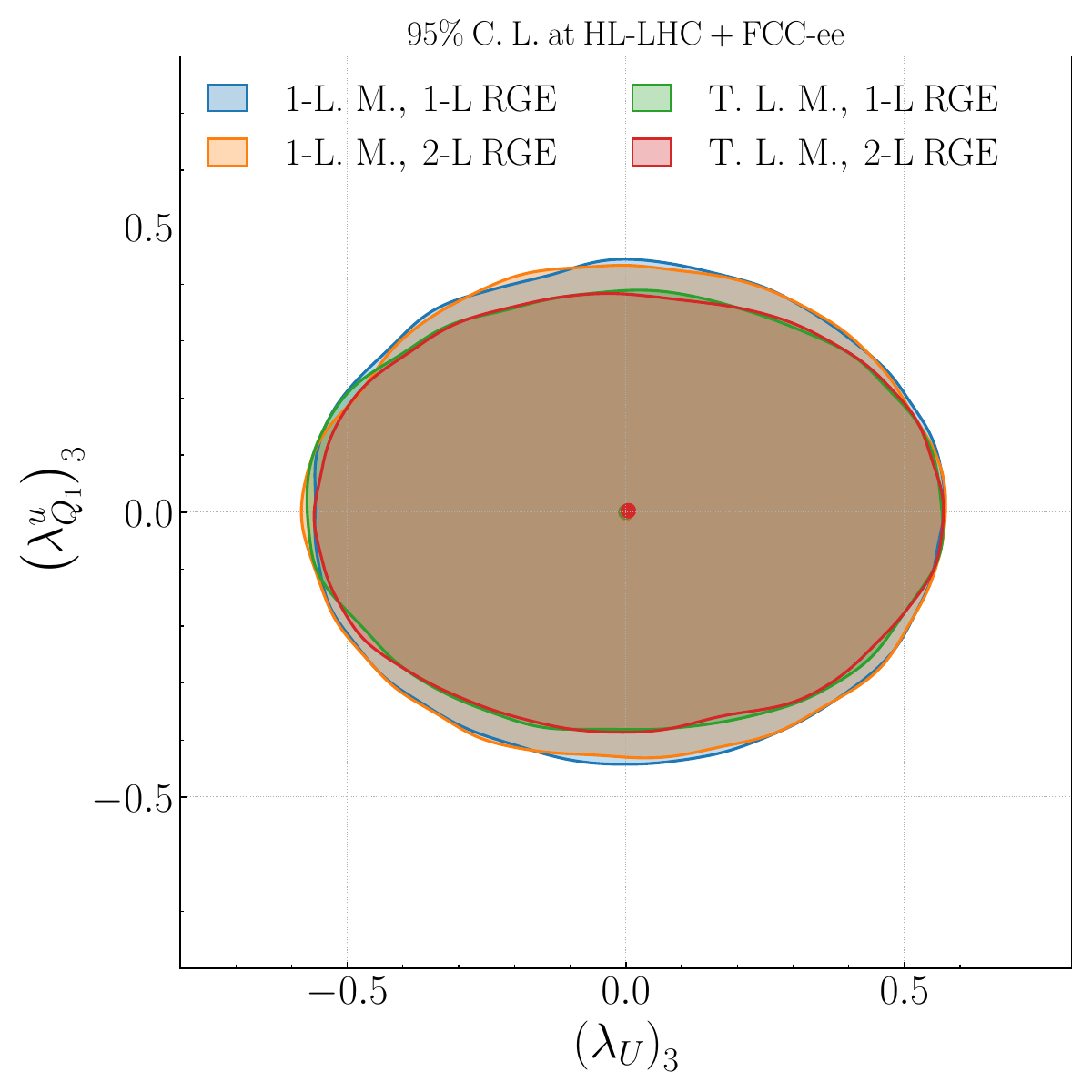}
    \caption{Two-dimensional $95\%$ confidence level region for the UV couplings $(\lambda_{Q_1}^{u})_3$ and $(\lambda_U)_3$ of the model that includes the vector-like fermions $U\sim(3,1)_{2/3}$ and $Q_{1}\sim(3,2)_{1/6}$, with masses $M_{U}=M_{Q_1}=10$~TeV, at HL-LHC (left) and HL-LHC+FCC-ee (right). 
    We consider tree-level (T. L. M.) and one-loop matching results (1-L. M.), and for each of them, we show the bounds with none, one-loop and two-loop SMEFT RGE effects. In the case of FCC-ee (right panel), we do not show the bounds without RGE effects since they include the whole plotted region.
    }
    \label{fig:2D_bounds_Q1plusU_HLLHC_FCCee}
\end{figure}

Figure~\ref{fig:2D_bounds_Q1plusU_HLLHC_FCCee} shows the $95\%$ credible regions 
in the $(\lambda_{Q_1}^{u})_{3}$--$(\lambda_U)_3$ plane at HL-LHC and FCC-ee. 
We find no significant improvement from two-loop RGEs over the one-loop result at either collider or 
matching level. At both colliders, the dominant gain comes from including one-loop 
RGEs, with one-loop matching reproducing only part of this improvement. At FCC-ee, 
the one-loop RGE improvement is so large that the $95\%$ credible region without 
running covers the entire plotted domain in both matching scenarios, and we 
therefore omit the no-running contour from the right panel of 
Fig.~\ref{fig:2D_bounds_Q1plusU_HLLHC_FCCee}.

When restricting our dataset to the ATLAS and 
CMS Run~2 Higgs inclusive signal strengths used in that work, we reproduce the result of~\cite{DiNoi:2025tka}
and recover a visible two-loop improvement in the tree-level matching 
scenario only. 
The richer dataset employed here introduces additional constraints that 
already saturate the sensitivity gain achievable through two-loop running, rendering 
the two-loop advantage unobservable.

\section{Conclusions}
\label{sec:conclusions}

The pursuit of New Physics through precision measurements demands theoretical 
predictions of matching accuracy. In the SMEFT framework, the program of 
higher-order calculations has been underway for over a decade, but only recently 
has the community crossed the frontier of two-loop computations, marked by the 
milestone of the complete two-loop SMEFT RGEs at dimension six.
In this work, we have presented the first systematic 
assessment of the phenomenological impact of this result.

Our approach proceeded in two steps. We first solved the two-loop RGE equations 
numerically and implemented the solution into the global fit framework \smefit, 
then used this to explore the phenomenology both bottom-up, through projected 
individual and global fits at HL-LHC and FCC-ee, and top-down, through a 
systematic study of the sensitivity to all scalar and fermionic one-particle 
extensions of the SM in the Granada dictionary consistent with our flavor symmetry assumption, matched to the SMEFT at one loop.

In individual fits, at both linear and quadratic level, the two-loop RGEs improve 
the sensitivity to several top-quark operators. The most notable effects are found 
for the dimension-6 top Yukawa operator and the color-octet two-light-two-heavy 
four-quark operators, driven by their two-loop-induced mixing into more tightly 
constrained directions. At quadratic level, the individual fits are largely unchanged: since most coefficients are well within the linear regime, the inclusion of quadratic EFT corrections has little impact on the comparison between one- and two-loop RGEs.

The global fits reveal a richer and more varied picture. The most striking effect 
is a loosening of the bound on $c_{\varphi G}$ by up to a factor of $\sim4$ in 
linear fits at the HL-LHC and by $50\%$ at FCC-ee, driven by the two-loop induced mixing of 
poorly constrained four-quark operators into $c_{\varphi G}$, which opens new 
correlations that dilute its bound. A qualitatively similar, though smaller, effect 
is observed for $c_{tG}$ at both colliders and for $c_{\varphi WB}$ at FCC-ee. 
In the opposite direction, the sensitivity to four-heavy quark operators is generally 
improved by the two-loop RGEs at HL-LHC, while a richer correlation pattern dilutes their bound at FCC-ee. 
Several of the effects present in the 
linear global fit are washed out once quadratic corrections are included, 
mirroring the known behaviour observed in one-loop RGE effects in global analyses~\cite{terHoeve:2025gey}.

The top-down study yields qualitatively consistent conclusions. Among the 
one-particle models of the Granada dictionary, two-loop RGEs improve the 
sensitivity to selected couplings by $2$--$5\%$, specifically in cases where those 
couplings generate significant contributions to $c_{t\varphi}$, the four-heavy-quark 
operators $c_{QQ}^{1}$ and $c_{QQ}^{8}$, or the two-lepton-two-quark operators 
$c_{Q\ell_3}^{(-)}$ and $c_{Q\ell_3}^{(3)}$, all of which benefit from enhanced 
two-loop mixing into tightly constrained bosonic operators. The dedicated analysis 
of the heavy scalar doublet confirms these findings at the percent level, with 
no qualitative difference observed between tree-level and one-loop matching. For 
the $Q_1+U$ model, by contrast, we find no significant advantage from two-loop 
RGEs at either matching level.

In addition, we performed fits to the quartic couplings in the models with scalars, whose effects only appear at one loop. Despite the loop suppression and assuming a new physics scale of $M=10$ TeV, we nevertheless find remarkable sensitivity to these couplings, often below the naive unitarity limit. This highlights the distinctive capability of FCC-ee to probe loop effects from heavy new physics.

Looking ahead, this work opens several directions for future exploration. The 
impact of two-loop RGEs on constraints from low-energy flavor observables could 
be substantial, given the rich new mixing patterns uncovered in 
Section~\ref{sec:comparison_1_vs_2_loops}, and deserves a dedicated study. The 
integration of the full two-loop RGE matrix without decoupling the SM parameter 
running could reveal additional features of relevance for flavor physics. On the 
model-building side, a broader exploration of UV completions where two-loop effects 
are maximized would be valuable, and the calculation of two-loop matching 
corrections for the models studied here could lift the suppression from SM 
couplings in the two-loop beta functions and sharpen the constraints on UV 
parameters. Ultimately, the full exploitation of these developments will require 
the computation of SMEFT observables at two-loop order. 
Finally, several of the phenomenological effects found here are due to $\gamma^5$- and evanescent-scheme dependent effects. Thus, in the spirit of~\cite{DiNoi:2025uhu}, it would be valuable to assess how a scheme change impacts the conclusions drawn in this work.

\section*{Acknowledgements}
We thank R. Gr\"ober, J. ter Hoeve, F. Maltoni and E. Vryonidou for useful discussions.
We are grateful to the authors of Ref.~\cite{DiNoi:2025tka} for giving us access to an updated version of their pre-print.
We thank R. Gr\"ober for her comments on an early version of this manuscript, and G. Ventura and E. Vryonidou for helping us to verify the basis rotation.
L.M. acknowledges support from the European Union under the MSCA fellowship (Grant agreement N. 101149078) {\it Advancing global SMEFT fits in the LHC precision era (EFT4ward)}.
The work of P.O. received funding from INFN Iniziative Specifiche APINE, from
the European Union's Horizon 2020 research and innovation programme under the Marie Sklodowska-Curie
grant agreement n. 101086085 - ASYMMETRY, and 
by the Italian MUR via the Departments of Excellence grant
2023- 2027 ``Quantum Frontiers'' and the PRIN 2022 project n. 2022K4B58X - AxionOrigins.
The work of A.N.R was funded by the University of Padua under the 2023 STARS Grants@UniPD Programme (Acronym and title of the project: HiggsPairs - Precise Theoretical Predictions for Higgs pair production at the LHC), by the Istituto Nazionale di Fisica Nucleare through its Iniziativa Specifiche APINE and RD-FCC, and by the Italian MUR via the Departments of Excellence grant
2023- 2027 ``Quantum Frontiers''.

\appendix
\newpage
\section{SMEFiT Basis}
\label{app:smefit_basis}

In this appendix, for completeness, we summarise the definitions of the SMEFT operators in the \smefit~basis.
For each operator, we indicate its definition in terms of the SM fields, and the conventions that are used both for the operator and for the coefficient. 
The predictions for all considered observables were computed assuming a $U(2)_{q_L}\times U(2)_{u_R}\times U(3)_{d_R}\times (U(1)_{\ell}\times U(1)_e)^3$ flavour symmetry or, in some cases, an even more general one.
Table~\ref{tab:oper_bos} lists the purely bosonic dimension-six SMEFT operators entering our basis.

\begin{table}[ht] 
  \begin{center}
    \renewcommand{\arraystretch}{1.5}
        \begin{tabular}{ll|ll}
          \toprule
          Coefficient & Definition & Coefficient & Definition \\
        \midrule
        $c_{\varphi G}$  & $\left(\pdp -\frac{v^2}{2} \right)G^{\mu\nu}_{\sss A}\,
        G_{\mu\nu}^{\sss A}$ 
        & 
        $c_{\varphi \square}$ & $(\pdp -\frac{v^2}{2} )\Box(\pdp)$ \\
         $c_{\varphi B}$ & $\left(\pdp -\frac{v^2}{2} \right)B^{\mu\nu}\,B_{\mu\nu}$
        &
         $c_{\varphi D}$ & $(\varphi^\dagger D^\mu\varphi)^\dagger(\varphi^\dagger D_\mu\varphi)$ \\ 
        $c_{\varphi W}$ & $\left(\pdp -\frac{v^2}{2} \right)W^{\mu\nu}_{\sss I}\,
        W_{\mu\nu}^{\sss I}$ 
        &
        $c_{WWW}$ & $\epsilon_{IJK}W_{\mu\nu}^I W^{J,\nu\rho} W^{K,\mu}_\rho$ \\ 
        $c_{\varphi W B}$ & $(\varphi^\dagger \tau_{\sss I}\varphi)\,B^{\mu\nu}W_{\mu\nu}^{\sss I}\,$ 
        & 
        $c_{\varphi}$ & $(\pdp -\frac{v^2}{2})^3$ \\
       \bottomrule
        \end{tabular}
        \vspace{0.2cm}
        \caption{The purely bosonic dimension-six SMEFT operators entering our basis.
        These operators modify the production and decay of Higgs bosons and the interactions of the electroweak gauge bosons.
        For each operator, we indicate its definition in terms of the SM
        fields,
        and the conventions that are used
        both for the operator and for the coefficient. 
        See~\cite{Ethier:2021bye} for more details.
        \label{tab:oper_bos}}
\end{center}
\end{table}

Table~\ref{tab:oper_ferm_bos} lists the operators containing two fermion fields, either quarks or leptons.
For the two-lepton operators, the flavour index $j$ runs from 1 to 3. 
Note that our basis includes the two-fermion operators modifying the Yukawa couplings of the top, bottom, and charm quarks ($c_{t\varphi}$, $c_{b\varphi}$, $c_{c\varphi}$) and of the tau and muon leptons ($c_{\tau\varphi}$, $c_{\mu\varphi}$).
The coefficients indicated with (*) in Table~\ref{tab:oper_ferm_bos} are replaced by  $c_{\varphi q}^{(-)}$, $c_{\varphi Q}^{(-)}$, and $c_{tZ}$.

\begin{table}[htbp]
  \begin{center}
    \renewcommand{\arraystretch}{1.25}
    \begin{tabular}{l l | l l}
      \toprule
       Coefficient & Definition &  Coefficient &  Definition \\
                \midrule
      \multicolumn{4}{c}{3rd-generation quarks} \\
                \midrule
     $c_{\varphi Q}^{(1)}$~(*) & $i\big(\varphi^\dagger\lra{D}_\mu\,\varphi\big)
 \big(\bar{Q}\,\gamma^\mu\,Q\big)$ 
 &
  $c_{tW}$ & $i\big(\bar{Q}\tau^{\mu\nu}\,\tau_{\sss I}\,t\big)\,
 \tilde{\varphi}\,W^I_{\mu\nu}
 + \text{h.c.}$ \\ 
     $c_{\varphi Q}^{(3)}$  & $i\big(\varphi^\dagger\lra{D}_\mu\,\tau_{\sss I}\varphi\big)
 \big(\bar{Q}\,\gamma^\mu\,\tau^{\sss I}Q\big)$ 
 &
 $c_{tB}$~(*) &
 $i\big(\bar{Q}\tau^{\mu\nu}\,t\big)
 \,\tilde{\varphi}\,B_{\mu\nu}
 + \text{h.c.}$\\ 
 $c_{\varphi Q}^{(-)}$ & $c_{\varphi Q}^{(1)}-c_{\varphi Q}^{(3)}$ 
 & $c_{tZ}$ & $-s_\theta \, c_{tB}+ c_\theta \,c_{tW}$ \\
 $c_{\varphi t}$& $i\big(\varphi^\dagger\,\lra{D}_\mu\,\,\varphi\big)
 \big(\bar{t}\,\gamma^\mu\,t\big)$
 & $c_{tG}$ & $ig{\sss S}\,\big(\bar{Q}\tau^{\mu\nu}\,T_{\sss A}\,t\big)\,
 \tilde{\varphi}\,G^A_{\mu\nu}
 + \text{h.c.}$ \\ 
 $c_{t\varphi}$ & $\left(\pdp\right)
 \bar{Q}\,t\,\tilde{\varphi} + \text{h.c.}$ 
 & $c_{b\varphi}$ & $\left(\pdp\right)
 \bar{Q}\,b\,\varphi + \text{h.c.}$ \\  
                \midrule
                \multicolumn{4}{c}{1st, 2nd generation quarks} \\
                \midrule
 $c_{\varphi q}^{(1)}$~(*) & $\sum\limits_{\sss i=1,2} i\big(\varphi^\dagger\lra{D}_\mu\,\varphi\big)
 \big(\bar{q}_i\,\gamma^\mu\,q_i\big)$ 
 &
 ${{c_{\varphi u}}}$ & $\sum\limits_{\sss i=1,2} i\big(\varphi^\dagger\,\lra{D}_\mu\,\,\varphi\big)
 \big(\bar{u}_i\,\gamma^\mu\,u_i\big)$ \\ 
 $c_{\varphi q}^{(3)}$ & $\sum\limits_{\sss i=1,2} i\big(\varphi^\dagger\lra{D}_\mu\,\tau_{\sss I}\varphi\big)
 \big(\bar{q}_i\,\gamma^\mu\,\tau^{\sss I}q_i\big)$
 &
      ${{c_{\varphi d}}}$ & $\sum\limits_{\sss j=1,2,3} i\big(\varphi^\dagger\,\lra{D}_\mu\,\,\varphi\big)
 \big(\bar{d}_j\,\gamma^\mu\,d_j\big)$ \\ 
 $c_{\varphi q}^{(-)}$ & $c_{\varphi q}^{(1)}-c_{\varphi q}^{(3)}$
 &
 $c_{c \varphi}$ & $\left(\pdp\right)
 \bar{q}_2\,c\,\tilde\varphi + \text{h.c.}$ \\ 
                \midrule
		      \multicolumn{4}{c}{Two leptons} \\
                \midrule
  $c_{\varphi \ell_j}$ & $ i\big(\varphi^\dagger\lra{D}_\mu\,\varphi\big)
   \big(\bar{\ell}_j\,\gamma^\mu\,\ell_j\big)$ 
   &
 $c_{\varphi e}$ & $ i\big(\varphi^\dagger\lra{D}_\mu\,\varphi\big)
 \big(\bar{e}\,\gamma^\mu\,e\big)$  \\
 $c_{\varphi \ell_j}^{(3)}$ & $ i\big(\varphi^\dagger\lra{D}_\mu\,\tau_{\sss I}\varphi\big)
 \big(\bar{\ell}_j\,\gamma^\mu\,\tau^{\sss I}\ell_j\big)$ 
 & $c_{\varphi \mu}$ & $ i\big(\varphi^\dagger\lra{D}_\mu\,\varphi\big)
 \big(\bar{\mu}\,\gamma^\mu\, \mu\big)$ \\  
 $c_{\mu \varphi}$ & $\left(\pdp\right)
 \bar{\ell_2}\,\mu\,{\varphi} + \text{h.c.}$ 
 & $c_{\varphi \tau}$ & $ i\big(\varphi^\dagger\lra{D}_\mu\,\varphi\big)
 \big(\bar{\tau}\,\gamma^\mu\,\tau\big)$  \\
  $c_{\tau \varphi}$ & $\left(\pdp\right)
 \bar{\ell_3}\,\tau\,{\varphi} + \text{h.c.}$ & & \\
  \bottomrule
\end{tabular}
\vspace{0.2cm}
\caption{Same as Table~\ref{tab:oper_bos}
  for the WCs containing two fermion fields, either quarks or leptons.
  Thy are classified in three groups: those involving 3rd generation quarks, those involving 1st and 2nd generation quarks, and those involving two leptons. 
  For the latter (two-lepton operators), the flavour index $j$ runs from 1 to 3 following our flavour assumptions. 
  The coefficients indicated with (*) do not correspond to physical degrees of freedom
  in the fit, but are rather replaced by  $c_{\varphi q}^{(-)}$, $c_{\varphi Q}^{(-)}$, and
  $c_{tZ}$.
\label{tab:oper_ferm_bos}}
\end{center}
\end{table}
\clearpage

Table~\ref{tab:oper_fourfermion} collects the definition of the relevant four-fermion WCs in terms of Warsaw basis coefficients.
These four fermion operators, involving either four-heavy quarks (4H), two-light-two-heavy quarks (2L2H), two-leptons-two-heavy quarks (2$\ell$2Q) or two-leptons-two-light quarks (2$\ell$2q), are defined as
\begin{align}
	\qq{1}{qq}{ijkl}
	&= (\bar q_i \gamma^\mu q_j)(\bar q_k\gamma_\mu q_l)
	 \nonumber
	,\\
	\qq{3}{qq}{ijkl}
	&= (\bar q_i \gamma^\mu \tau^I q_j)(\bar q_k\gamma_\mu \tau^I q_l)
     \nonumber
	,\\
	\qq{1}{qu}{ijkl}
	&= (\bar q_i \gamma^\mu q_j)(\bar u_k\gamma_\mu u_l)
     \nonumber
	,\\
	\qq{8}{qu}{ijkl}
	&= (\bar q_i \gamma^\mu T^A q_j)(\bar u_k\gamma_\mu T^A u_l)
         \nonumber
	,\\
	\qq{1}{qd}{ijkl}
	&= (\bar q_i \gamma^\mu q_j)(\bar d_k\gamma_\mu d_l)
	,\\
	\qq{8}{qd}{ijkl}
	&= (\bar q_i \gamma^\mu T^A q_j)(\bar d_k\gamma_\mu T^A d_l)
        \label{eq:4f_def}
        \nonumber
	,\\
	\qq{}{uu}{ijkl}
	&=(\bar u_i\gamma^\mu u_j)(\bar u_k\gamma_\mu u_l)
         \nonumber
	,\\
	\qq{1}{ud}{ijkl}
	&=(\bar u_i\gamma^\mu u_j)(\bar d_k\gamma_\mu d_l)
         \nonumber
	,\\
	\qq{8}{ud}{ijkl}
	&=(\bar u_i\gamma^\mu T^A u_j)(\bar d_k\gamma_\mu T^A d_l) \, .
         \nonumber 
\end{align}
Moreover, we include the four-lepton (4$\ell$) operators
\begin{align}
\nonumber
    \qq{}{\ell \ell}{jklm}
    &= \left(\bar \ell_j\gamma_\mu \ell_k\right) \left(\bar \ell_l\gamma^\mu \ell_m\right) 
    ,\\
    \qq{}{\ell e}{jklm}
    &= \left(\bar \ell_j\gamma_\mu \ell_k\right) \left(\bar e_l\gamma^\mu e_m\right)
    ,\\
    \qq{}{e e}{jklm}
    &= \left(\bar e_j\gamma_\mu e_k\right) \left(\bar e_l\gamma^\mu e_m\right)
    \, .
         \nonumber
\end{align}
The flavour index $i$ is either 1 or 2, and $j$ is either 1, 2 or 3.

\begin{table}[htbp] 
  \begin{center}
    \renewcommand{\arraystretch}{1.53}
        \begin{tabular}{ll| ll}
          \toprule
          \smefit &  Warsaw  & \smefit &  Warsaw \\
          \midrule
          \multicolumn{4}{c}{4-heavy-quark operators} \\
          \midrule
      $c_{QQ}^1$    &   $2\ccc{1}{qq}{3333}-\frac{2}{3}\ccc{3}{qq}{3333}$ 
      &
      $c_{QQ}^8$       &         $8\ccc{3}{qq}{3333}$\\  
     $c_{Qt}^1$         &         $\ccc{1}{qu}{3333}$
     &
     $c_{Qt}^8$         &         $\ccc{8}{qu}{3333}$\\   
     $c_{tt}^1$         &         $\ccc{1}{uu}{3333}$ && \\
          \midrule
            \multicolumn{4}{c}{2-light-2-heavy quark operators} \\
          \midrule
  $c_{Qq}^{1,8}$       &  	 $\ccc{1}{qq}{i33i}+3\ccc{3}{qq}{i33i}$  
  &
  $c_{Qq}^{1,1}$         &   $\ccc{1}{qq}{ii33}+\frac{1}{6}\ccc{1}{qq}{i33i}+\frac{1}{2}\ccc{3}{qq}{i33i} $   \\    
   $c_{Qq}^{3,8}$         &   $\ccc{1}{qq}{i33i}-\ccc{3}{qq}{i33i} $  
   &
$c_{Qq}^{3,1}$          & 	$\ccc{3}{qq}{ii33}+\frac{1}{6}(\ccc{1}{qq}{i33i}-\ccc{3}{qq}{i33i}) $   \\     
$c_{tq}^{8}$         &  $ \ccc{8}{qu}{ii33}   $ 
  &
$c_{tq}^{1}$       &   $  \ccc{1}{qu}{ii33} $\\   
$c_{tu}^{8}$      &   $2\ccc{}{uu}{i33i}$ 
 &
$c_{tu}^{1}$        &   $ \ccc{}{uu}{ii33} +\frac{1}{3} \ccc{}{uu}{i33i} $ \\   
$c_{Qu}^{8}$         &  $  \ccc{8}{qu}{33ii}$
 &
 $c_{Qu}^{1}$     &  $  \ccc{1}{qu}{33ii}$  \\    
 $c_{td}^{8}$        &   $\ccc{8}{ud}{33jj}$ 
  &
 $c_{td}^{1}$          &  $ \ccc{1}{ud}{33jj}$ \\    
 $c_{Qd}^{8}$        &   $ \ccc{8}{qd}{33jj}$ 
 &
 $c_{Qd}^{1}$         &   $ \ccc{1}{qd}{33jj}$\\
 
          \midrule  
              \multicolumn{4}{c}{2-heavy-quark-2-lepton operators} \\
          \midrule
  $c_{Q\ell_1}^{(-)}$       &  	 $\ccc{1}{\ell q}{1133}-\ccc{3}{\ell q}{1133}$  
  &
  $c_{Q\ell_1}^{(3)}$         &   $\ccc{3}{\ell q}{1133}$   \\    
  $c_{Q\ell_3}^{(-)}$       &  	 $\ccc{1}{\ell q}{3333}-\ccc{3}{\ell q}{3333}$  
  &
  $c_{Q\ell_3}^{(3)}$         &   $\ccc{3}{\ell q}{3333}$   \\  
  $c_{Q e}$       &  	 $\ccc{}{qe}{3311}$  
  & $c_{Q \tau}$       &  	 $\ccc{}{qe}{3333}$ \\ 
  $c_{t e}$       &  	 $\ccc{}{eu}{1133}$  & $c_{t \tau}$       &  	 $\ccc{}{eu}{3333}$   \\
  $c_{t\ell_j}$         &   $\ccc{}{\ell u}{jj33} $  & \\        
  \midrule  
  \multicolumn{4}{c}{2-light-quark-2-lepton operators} \\
  \midrule
  $c_{q\ell_1}^{(-)}$       &  	 $\ccc{1}{\ell q}{11ii}-\ccc{3}{\ell q}{11ii}$  
  &
  $c_{q\ell_1}^{(3)}$         &   $\ccc{3}{\ell q}{11ii}$   \\    
  $c_{q e}$       &  	 $\ccc{}{qe}{ii11}$   & & \\
  $c_{\ell_1 u}$         &   $\ccc{}{\ell u}{11ii} $   &  
  $c_{\ell_1 d}$         &   $\ccc{}{\ell d}{11jj} $   \\    
  $c_{e u}$         &   $\ccc{}{e u}{11ii} $   &  
  $c_{e d}$         &   $\ccc{}{e d}{11jj} $   \\ 
  \bottomrule
  \end{tabular}
  \vspace{0.2cm}
  \caption{\small Definition of the four-fermion coefficients that in
    the \smefit~basis in terms of Warsaw basis coefficients, with $i=1,\,2$ and $j=1,\,2,\,3$. 
  \label{tab:oper_fourfermion}}
  \end{center}
\end{table}

In the global fits presented in Sect.~\ref{sec:smeft_fits} we impose a $U(3)_{\ell}\times U(3)_e$ symmetry in the lepton sector.
The flavour universal coefficients definitions are trivially derived by identifying all lepton flavour indices.
The four-lepton operator $\qq{}{\ell \ell}{jklm}$, is split into two invariants, such that~\cite{Brivio:2020onw}
\begin{align}
    c_{\ell \ell}^{jklm} = c_{\ell \ell}\delta^{jk}\delta^{lm} + c_{\ell \ell}^{\prime}\delta^{jm}\delta^{lk} \, .
\end{align}

\clearpage
\section{Reduced fit illustrating two-loop degeneracy breaking}
\label{app:red_fit}

To isolate and illustrate the mechanism responsible for the improved sensitivity 
to $c_{\varphi D}$ and $c_{\varphi t}$ discussed in Section~\ref{sec:global_fits}, 
we construct a reduced fit designed to amplify the interplay between four-heavy-quark 
operators and the bosonic sector induced by two-loop RGEs.

The dataset used in this fit comprises three complementary sets of observables: 
Electroweak Precision Observables, Higgs signal strength 
projections at the HL-LHC and four-heavy-quark production measurements at the HL-LHC, specifically 
four-top and $t\bar{t}b\bar{b}$ production, which directly constrain the 
four-heavy-quark sector. The operator basis consists of seven Wilson coefficients: 
$c_{\varphi D}$, $c_{\varphi t}$, and the five four-heavy-quark operators 
$c_{QQ}^{1}$, $c_{QQ}^{8}$, $c_{Qt}^{1}$, $c_{Qt}^{8}$, and $c_{tt}^{1}$.

\begin{figure}[h!]
    \centering
    \includegraphics[width=\linewidth]{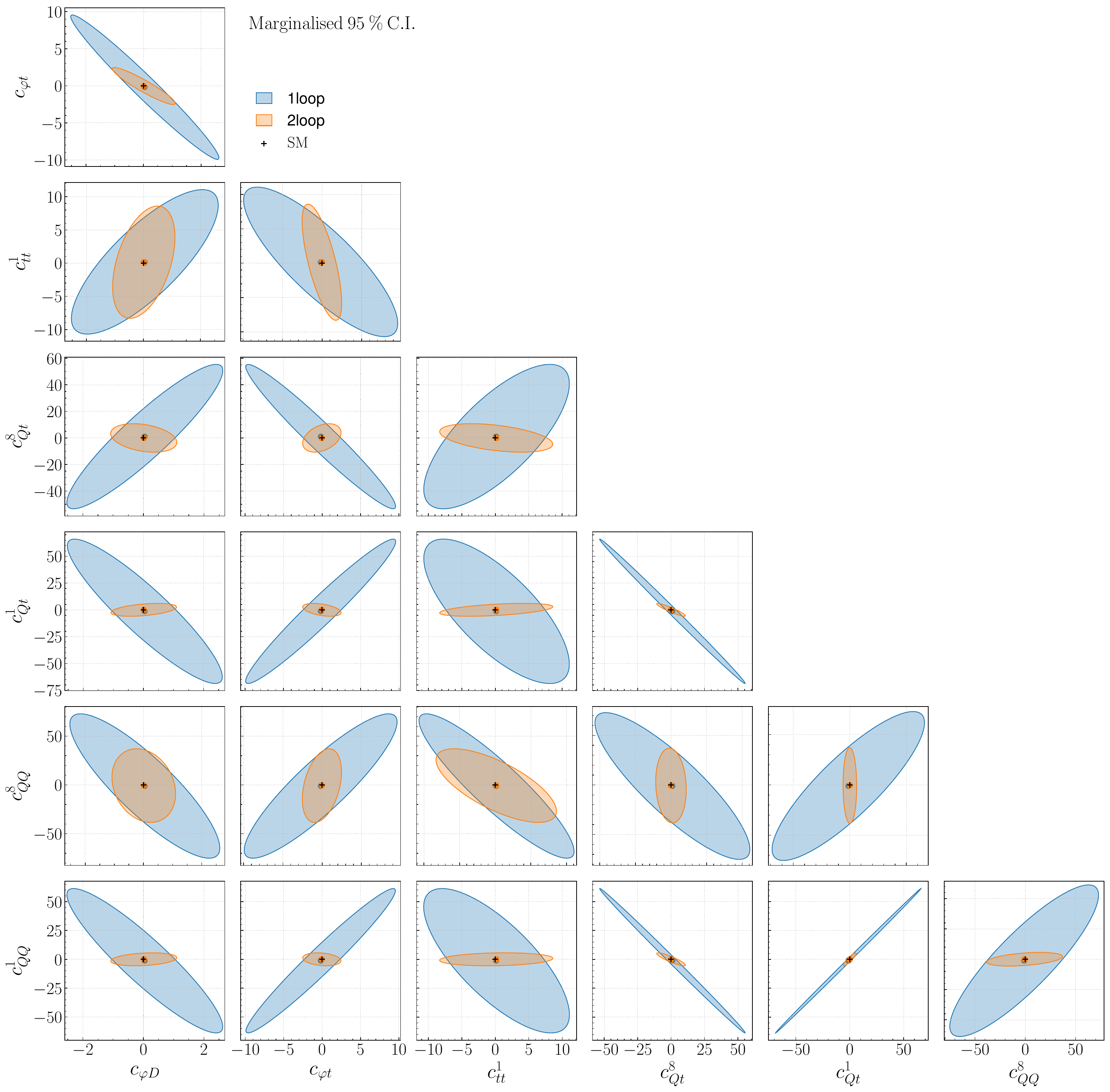}
    \caption{Pairwise $95\%$ credible regions for all pairs of fitted operators 
    in the reduced fit of Appendix~\ref{app:red_fit}, comparing one-loop (blue) 
    and two-loop (red) SMEFT RGEs. The fitted operators are $c_{\varphi D}$, 
    $c_{\varphi t}$, $c_{QQ}^{1}$, $c_{QQ}^{8}$, $c_{Qt}^{1}$, $c_{Qt}^{8}$, 
    and $c_{tt}^{1}$. The dataset comprises EWPOs, Higgs signal strength 
    projections, and four-top and $t\bar{t}b\bar{b}$ production measurements 
    at the HL-LHC.}
    \label{fig:corner_reduced_fit}
\end{figure}

\begin{figure}[h!]
    \centering
    \includegraphics[width=0.8\linewidth]{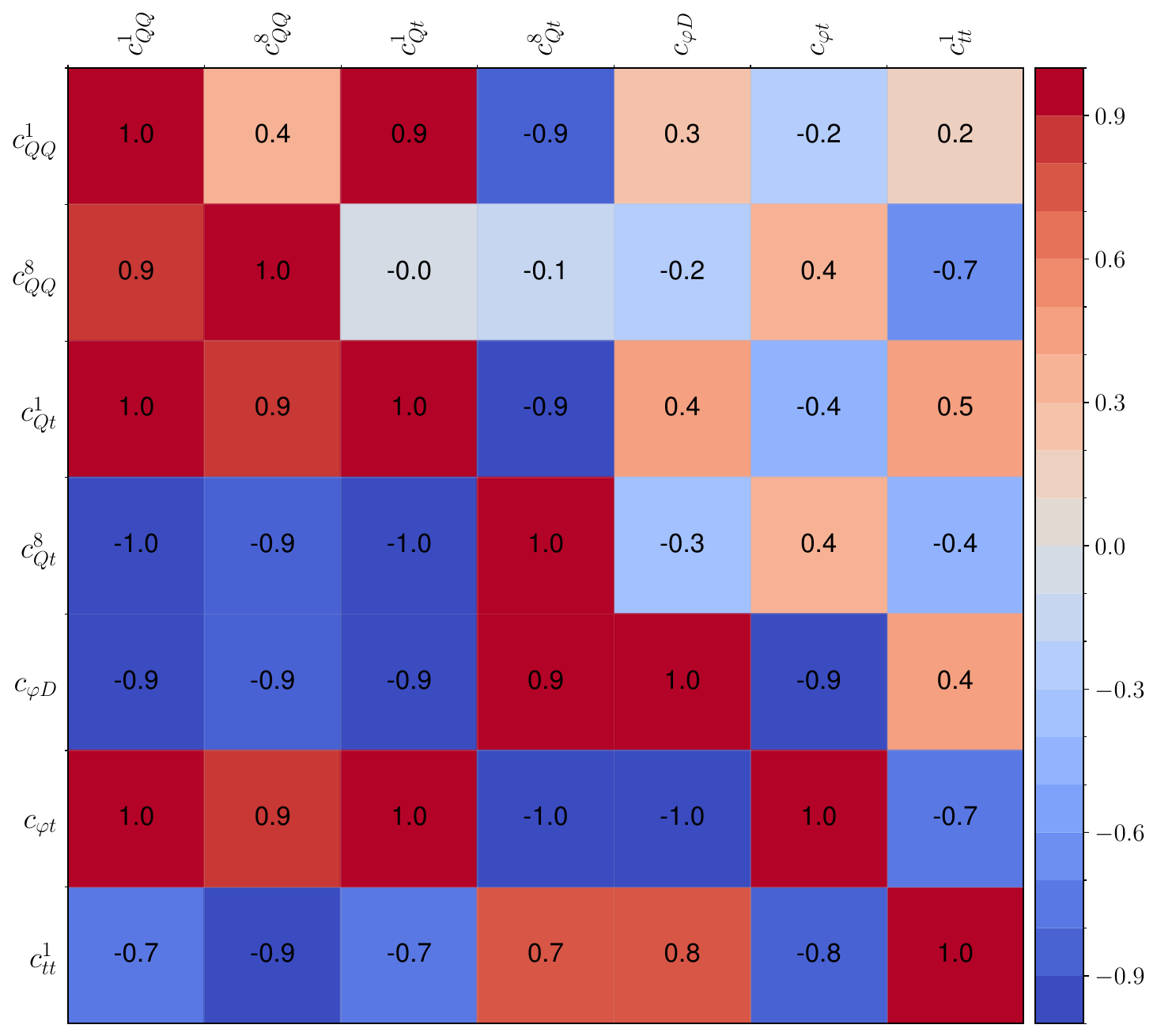}
    \caption{Correlation matrix for the reduced fit of Appendix~\ref{app:red_fit}, 
    with one-loop RGE correlations shown in the lower triangle and two-loop RGE 
    correlations in the upper triangle. The fitted operators and dataset are as 
    described in Fig.~\ref{fig:corner_reduced_fit}.}
    \label{fig:corr_reduced_fit}
\end{figure}

We perform two fits, at linear order, identical in all respects except for the loop order of the 
SMEFT RGEs: one with one-loop running and one with two-loop running. The results 
are shown in Figs.~\ref{fig:corner_reduced_fit} and~\ref{fig:corr_reduced_fit}. 
Figure~\ref{fig:corner_reduced_fit} displays the pairwise $95\%$ credible regions 
for all operator pairs, while Figure~\ref{fig:corr_reduced_fit} shows the 
correlation matrices, with one-loop correlations in the lower triangle and 
two-loop correlations in the upper triangle.

The comparison makes the degeneracy-breaking mechanism transparent. At one loop, 
all operators in the fit derive their sensitivity predominantly from EWPOs, leading 
to strong correlations among them and diluting the constraints on each individual 
operator. At two loops, the running of the four-heavy-quark operators into Higgs 
production observables introduces an independent source of information on the 
four-heavy-quark sector, breaking the degeneracies with $c_{\varphi D}$ and 
$c_{\varphi t}$ and allowing the different datasets to constrain orthogonal 
directions in parameter space. The net effect is a tightening of the bounds on 
all operators in the fit, here magnified with respect to the full global fit by 
the reduced operator basis and dataset.

\clearpage

\section{Details on the UV models}
\label{app:UV_details}

In Table~\ref{tab:UV_models_and_couplings} we list all the UV couplings for each model that appear in the one-loop matching results relevant to our operator basis.
The normalization and naming of the couplings that enter only at one-loop level follows the one of \sold~\cite{Guedes:2023azv,Guedes:2024vuf}. For the other couplings, the naming and normalization follows the Granada dictionary~\cite{deBlas:2017xtg}.
As said in the main text, we assume that the heavy particles interact with the bosonic and third-generation fermionic sectors of the SM.

\begin{table}[hbp]
    \centering
    \begin{tabular}{c|c || c | c}
        Scalar & Couplings & Fermion & Couplings \\
        \hline
        $S$ & $\kappa_S$, $\kappa_{S^3}$, $\lambda_{S}$  $L_{S}^{(3)}$ & $N$ & $(\lambda_N)_3$ \\
        $S_1$ & {\color{gray}$(y_{S_1})_{12}$, $(y_{S_1})_{21}$,} $L_{S_1}^{(2)}$, $L_{S_1}^{(3)}$ & $E$ & $(\lambda_E)_3$ \\
        $S_2$ &  $(y_{S_2})_{33}${\color{gray}, $L_{S_2}^{(1)}$, $L_{S_2}^{(2)}$} & $\Delta_1$ & $(\lambda_{\Delta_1})_3$ \\
        $\phi$ & $(y_{\phi}^{u})_{33}$, $\lambda_\phi$, $L_{\phi}^{(2)}$, $L_{\phi}^{(3)}$, $L_{\phi}^{(4)}$  & $\Delta_3$ & $(\lambda_{\Delta_3})_3$ \\
        $\Xi$ & $\kappa_{\Xi}$, $\lambda_{\Xi}${\color{gray}, 
        $L_{\Xi}^{(2)}$, $L_{\Xi}^{(3)}$ }  & $\Sigma$ & $(\lambda_{\Sigma})_3$ \\
        $\Xi_1$ & $\kappa_{\Xi_1}$, 
        $L_{\Xi_1}^{(2)}$, {\color{gray} $L_{\Xi_1}^{(4)}$,} $L_{\Xi_1}^{(5)}$, $(L_{\Xi_1}^{(6)})_{33}$  & $\Sigma_1$ & $(\lambda_{\Sigma_1})_3$ \\
        $\Theta_1$ & $\lambda_{\Theta_1}$, $L_{\Theta_1}^{(2)}$, {\color{gray}$L_{\Theta_1}^{(3)}$,} $L_{\Theta_1}^{(4)}${\color{gray}, $L_{\Theta_1}^{(7)}$, $L_{\Theta_1}^{(8)}$} & $U$ & $(\lambda_U)_3$ \\
        $\Theta_3$ & $\lambda_{\Theta_3}$, {\color{gray}$L_{\Theta_3}^{(2)}$,} $L_{\Theta_3}^{(3)}${\color{gray}, $L_{\Theta_3}^{(4)}$, $L_{\Theta_3}^{(5)}$} & $D$ & $(\lambda_D)_3$ \\
        $\omega_1$ & $(y_{\omega_1}^{qq})_{33} $, $(y_{\omega_1}^{q\ell})_{33} $, 
        $L_{\omega_1}^{(1)} $, $L_{\omega_1}^{(2)}$  & $Q_1$ & $(\lambda_{Q_1})_3$ \\
        $\omega_4$ & $L_{\omega_4}^{(1)}$ & $Q_7$ & $(\lambda_{Q_7})_3$ \\
        $\Pi_1$ & $L_{\Pi_1}^{(2)}$, $L_{\Pi_1}^{(3)}$  & $T_1 $ & $(\lambda_{T_1})_{3}$ \\
        $\Pi_7$ & $(y_{\Pi_7}^{\ell u})_{33}$, $L_{\Pi_7}^{(1)}$, $L_{\Pi_7}^{(2)}$, $L_{\Pi_7}^{(3)}${\color{gray}, $L_{\Pi_7}^{(4)}$}  & $T_2 $ & $(\lambda_{T_2})_{3}$ \\
        $\zeta$ & $(y_{\zeta}^{q\ell})_{33}$, {\color{gray} $L_{\zeta}^{(1)}$, }$L_{\zeta}^{(2)}${\color{gray}, $L_{\zeta}^{(3)}$, $L_{\zeta}^{(4)}$, $L_{\zeta}^{(5)}$}   &  \\
        $\Omega_1$ & $(y_{\Omega_1}^{qq})_{33}$, $L_{\Omega_1}^{(1)}$, \color{gray}{$L_{\Omega_1}^{(2)}$, $L_{\Omega_1}^{(3)}$}   & & \\
        $\Omega_2$ & $L_{\Omega_2}^{(1)}$ & & \\
        $\Omega_4$ & $(y_{\Omega_4}^{uu})_{33}$, \color{gray}{$L_{\Omega_4}^{(1)}$, $L_{\Omega_4}^{(2)}$}, , $L_{\Omega_4}^{(3)}$   & & \\
        $\Upsilon$ & $(y_\Upsilon)_{33}$, $L_\Upsilon^{(1)}$, $L_\Upsilon^{(2)}$ &  & \\
        $\Phi$ & $(y_{\Phi}^{qu})_{33}$, $L_\Phi^{(2)}$, $L_\Phi^{(3)}$ & &
    \end{tabular}
    \caption{List of UV models and their couplings considered in our analysis. We follow the notation of the Granada dictionary and, for couplings that enter only via one-loop matching contributions, we adopt the \sold~notation. Couplings shown in grey contribute to the matching results, but to which we have no significant sensitivity and, therefore, we marginalise over them with a flat prior in $[-15,\,15]$. For dimensionful couplings, we set them in units of TeV.}
    \label{tab:UV_models_and_couplings}
\end{table}

A few models present in the Granada dictionary have not been included in our study.         $\omega_2$ and $\Omega_2$ couple to down-type quarks via the operator $\bar{d}^{A}_{Ri} {d^c_{R}}^{B}_{j}$, with $A,\,B$ being color indices, and hence cannot generate any U$(3)_d$ symmetric dimension-6 interaction. The heavy state $Q_5$ interacts with down-type quarks and generates tree-level contributions to $\mcO_{\varphi d}$ and $\mcO_{d\varphi}$. However, it is incapable of generating them in a U$(3)_d$ symmetric manner: requiring that the non-diagonal contributions to $\mcO_{\varphi d}$ vanish leads to vanishing UV couplings.

\clearpage

\bibliographystyle{JHEP}
\bibliography{main}

\end{document}